\documentclass{article}
\usepackage[utf8]{inputenc}

\usepackage[T1]{fontenc}
\usepackage{graphicx}

\usepackage{epstopdf}
\usepackage{amsmath}
\usepackage{dsfont}

\usepackage{cite}

\usepackage{booktabs}

\usepackage{url}

\usepackage{footnote}
\makesavenoteenv{tabular}
\usepackage{comment}

\usepackage{geometry}
\usepackage{pdflscape}
\usepackage{afterpage}

\usepackage{color}   %May be necessary if you want to color links
\usepackage{hyperref}
\hypersetup{
    colorlinks=true, %set true if you want colored links
    linktoc=all,     %set to all if you want both sections and subsections linked
    linkcolor=blue,  %choose some color if you want links to stand out
}
\hypersetup{linktocpage}

\title{Biology-inspired geometric representation of probability and applications to completion and options' pricing}

\author{Felix Polyakov \\
\includegraphics[scale=0.65]{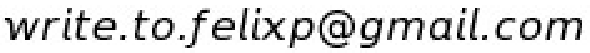} \\  \\
 Department of Mathematics\\
 Bar Ilan University\\
 Ramat-Gan, Israel}
\date{\today} % March 2018}

\usepackage{amsfonts}
\usepackage{cases}

\usepackage{eurosym}

\begin{document}

\maketitle

\begin{abstract}
\par Geometry constitutes a core set of intuitions present in all humans, regardless of their language or schooling \cite{dehaene.izard.pica.spelke:2006}. Could brain's built in machinery for processing geometric information take part in uncertainty representation? For decades already traders have been citing the price of uncertainty based FX optional contracts in terms of implied volatility, a dummy variable  related to the standard deviation, instead of pricing with units of money. This work introduces a methodology for geometric representation of probability in terms of implied volatility and attempts to find ways to approximate certain probability distributions using intuitive geometric symmetry.
\par In particular, it is shown how any probability distribution supported on $\mathbb{R}_{+}$ and having finite expectation may be represented with a planar curve whose geometric characteristics can be further analyzed. Log-normal distributions are represented with circles centered at the origin. Certain non-log-normal distributions with bell-shaped density profiles are represented by curves that can be closely approximated with circles whose centers are translated away from the origin. Only three points are needed to define a circle while it represents the candidate probability density approximating the distribution along the entire $\mathbb{R}_{+}$. Just three numbers: scaling and translations along the $x$ and $y$ axes map one circle to another. It is possible to introduce equivalence classes whose member distributions can be obtained by transitive actions of geometric transformations on any of corresponding representations.
\par Approximate completion of probability with non-circular shapes and cases when probability is supported outside of $\mathbb{R}_{+}$ are considered too. Proposed completion of implied volatility is compared to the vanna-volga method.
\end{abstract}

\tableofcontents

\section{Motivation}
\par Geometric mechanisms constitute an essential part of conscious and subconscious cognitive processes of humans, e.g. \cite{dehaene.izard.pica.spelke:2006, Viviani_Schneider_1991}. Geometric representation of uncertainty proposed here leads to a simple and intuitive procedure for approximation of several common families of probability distributions. Presumably, evolution causes biological systems to optimize the computational load of their performance; therefore, representation of uncertainty in human cognition may follow computational sub-optimality. Brain's built in ability to process geometric information is utilized for visual, motor and other cognitive tasks; its involvement into processing uncertainty may be computationally beneficial to reduce recruitment of additional computational circuits. The above biological rationale is mainly conceptual and proposes perspectives for empirical and computational studies.
\par The proposed methodology relies on both:
\begin{enumerate}
  \item The way the uncertainty is priced by human professionals: options' traders. Traders measure uncertainty with implied volatility $\sigma$ that, roughly speaking, stands for the $\sqrt{\mbox{time}}$-scaled standard deviation of $\log(\mbox{Uncertain future asset price})$. Correct value of implied volatility $\sigma(K)$ being plugged into the Black-Scholes-Merton formula\footnote{Used for pricing future difference between the future asset price the value of $K$.} results in the market price of vanilla option with strike $K$.
  \item Intuitive geometric concepts, for instance translation in plane.
\end{enumerate}
\par Geometric visualization generally helps to get an intuitive understanding of different mathematical concepts and to find solutions to problems. For example, an effective teaching of fractions is based on ``cutting a pie''\footnote{Henri Poincar\'e said \cite{arnol'd:2006}: "If you want to teach fractions either you divide cakes, even in a virtual way or you bring an apple to the classroom. In all other cases students will continue adding numerators and adding denominators".}, normal distribution with two variables can be identified with ellipse, integral of a scalar function is identified with area under a curve; geometric symmetries allow to solve or simplify certain differential equations \cite{Olver:1993, Ibragimov:1999}, and so on.
\par The proposed approach establishes geometric connection among different probability distributions. For example, translation and scaling connect among distributions represented with circles that in turn may presumably provide close approximation to distributions with bell-shaped density profiles. More than that, geometric representation and visualization of probability distributions would allow to employ geometric tools and geometric intuition into probabilistic reasoning.

\section{Background}
\par The ``Background'' section contains well known facts from financial mathematics and geometry. Books \cite{hull_book, Haug:2007, Shirokovy_1959, OlsenJ:2010} and numerous other works can serve as reference. The variables from financial mathematics employed in this work and their meaning are introduced in Table \ref{tab:meaning.variable.callprice}.
\subsection{Risk-neutral probability density and vanilla options' prices}
\par Let $p(S_T)$ be probability density of the random variable $S_T$ with expectation $\mathbb{E}(S_T)$ and $P(S_T)$ be the corresponding cumulative density function. Let $r$, $q$ be arbitrary real numbers and $T$ an arbitrary non-negative number. Denote
\begin{equation}\label{eq:S0.expectation}
  S_0 = e^{-(r - q) T} \cdot \mathbb{E}(S_T)
\end{equation}
and
\begin{eqnarray}\label{eq:callprice}
    \mbox{call}(K) \equiv e^{-r T} \mathbb{E}\left([S_T - K]^{+}\right) = e^{-r T} \int_{K}^{\infty}\left(S_{T} - K \right) \cdot p(S_T) d S_T \, , \\
    \label{eq:putprice}
    \mbox{put}(K) \equiv e^{-r T} \mathbb{E}\left([K - S_T]^{+}\right) = e^{-r T} \int_{-\infty}^{K}\left(K - S_{T} \right) \cdot p(S_T) d S_T \, .
\end{eqnarray}

Consequently,
\begin{equation}\nonumber %\label{eq:1st.derivative.call}
    \displaystyle\frac{\partial \mbox{call}(K)}{\partial K} = -e^{-r T} \int_{K}^{\infty}p(S_T) d S_T = -e^{-r T} \left[1 - P(K)\right] = e^{-r T} \left[ P(K) - 1 \right]\,
\end{equation}
and
\begin{equation}\label{eq:2nd.derivative.call}
    \displaystyle\frac{\partial^2 \mbox{call}(K)}{\partial K^2} = e^{-r T} p(K)\, .
\end{equation}

Equations \eqref{eq:S0.expectation}, \eqref{eq:callprice}, \eqref{eq:putprice}
imply the put-call parity
\begin{equation}\nonumber%\label{eq:put.call.parity}
    \mbox{call}(K) - \mbox{put}(K) = e^{-r T} \int_{-\infty}^{\infty}\left(S_{T} - K \right) \cdot p(S_T) d S_T = e^{-q T} S_0 - e^{-r T} K\, .
\end{equation}

\begin{table}[hbt!]
\caption{Notation.}
\begin{tabular}{|p{0.1\linewidth}|p{0.75\linewidth}|}%{|c|l|}
    \toprule
    \multicolumn{1}{|c|}{Variable} & \multicolumn{1}{|c|}{Definition or meaning in financial mathematics}\\
      \bottomrule
  $S_0$          &  Spot: the currently traded value of an asset like exchange rate or a stock price \\ \hline
  $r$          & Interest rate of the domestic currency: if the price of the traded asset is measured in USD, then USD interest for the period $T$ \\ \hline
  $q$          & Interest rate of the foreign currency or dividend yield of the stock. Eg EUR interest for EUR/USD underlying \\ \hline
  call          &   Price of European vanilla call option (the right to buy an asset in future time $T$ for the predefined strike value K)  \\ \hline
   put          &   Price of European vanilla put option (the right to sell an asset in future time $T$ for the predefined strike value K)  \\ \hline
   $K$ & Strike value of a financial derivative \\ \hline
  $T$          & Time left to the moment when European option can be exercised (time to option's expiry) \\ \hline
  $S_T$      & Price of asset $S$ at future time $T$ \\ \hline
  $p(S_T)$        & Risk-neutral/implied probability density of future asset's price $S_T$.
  \par {\footnotesize The word ``implied'' will sometimes be omitted further in text. Computations of volatility smiles in this work are related exclusively to implied probabilities. The term ``implied'' is used as values of $p$ and $\sigma$ may be numerically implied from vanilla option prices known continuously over $K$.} \\ \hline
  $P(S_T)$        & Cummulative implied probability density  of future asset's price $S_T$ \\ \hline
  $\sigma(K)$      & \emph{Implied volatility}; for any given $T$, when the value of $\sigma$ is the same for any strike $K$, then $p(S_T)$ is log-normal and $\sigma$ is equal to the standard deviation of annualized continuous returns of $S_T$ (standard deviation of $\ln\left(S_T / S_0\right) / \sqrt{T}$) \\  \hline
  $d_1(K)$ &     $$\frac{\ln(S_0 / K) + \left(r - q + \sigma^2 / 2 \right) T}{\sigma \sqrt{T}}$$  \\ \hline
  $d_2(K)$ & $$\frac{\ln(S_0 / K) + \left(r - q - \sigma^2 / 2 \right) T}{\sigma \sqrt{T}} = d_1(K) - \sigma(K) \sqrt{T}$$ \\ \hline
  $N(x)$ & $$\frac{1}{\sqrt{2 \pi}} \int_{-\infty}^{x}   e^{-y^2/2} dy $$ \\ \hline
  $n(x)$ & $$N'(x) =  \frac{1}{\sqrt{2 \pi}} e^{-x^2/2} $$  \\ \hline
    $\Delta_{\mbox{call}}$ &
  $ \partial \mbox{call} / \partial S_0$
  \par In case $S_T$ is log-normally distributed, $\Delta_{\mbox{call}} = e^{-q T} N(d_1)$ \\ \hline
  $\Delta_{\mbox{put}}$ &
  $ \partial \mbox{put} / \partial S_0 = \Delta_{\mbox{call}} - e^{-q T}$
  %%\par When $\sigma \sqrt{T}$ is relatively small, $\Delta_{\mbox{put}}$
  In case $S_T$ is log-normally distributed, $\Delta_{\mbox{put}} -e^{-q T} N(-d_1)$.
  \par Is used by option traders to represent strikes of cited options. For example, for $T < 1$ and log-normally distributed $S_T$, ``25 delta put'' stands for such strike $K_{25}$ that $ |\Delta_{\mbox{put}}(K_{25})| = e^{-q T} N(-d_1(K_{25})) = 0.25 = 25\%$ \\ \hline
  $\mbox{ATMF}$ & At the money forward (strike) $\equiv \mathbb{E}(S_T) = e^{(r - q) T} \cdot S_0$
  \\ \hline
  $\mbox{ATM}_{\mbox{rn}}$ & At The Money risk neutral strike that zeros the delta of the portfolio $\mbox{call} + \mbox{put}$: $\Delta_{\mbox{call}}(\mbox{ATM}_{\mbox{rn}}) + \Delta_{\mbox{put}}(\mbox{ATM}_{\mbox{rn}}) = 0$.
  \par For log-normally distributed $S_T$, $\mbox{ATM}_{\mbox{rn}} = S_0 e^{(r - q + \sigma^2/2) T}$ \\ \hline
  \end{tabular}
\label{tab:meaning.variable.callprice}
\end{table}

\par The values of call and put from formulae \eqref{eq:callprice}, \eqref{eq:putprice} are usually represented in terms of a dummy variable $\sigma(K)$ called \emph{implied volatility} by means of the Black-Scholes-Merton (BSM) formula:
\begin{equation}\label{eq:BSM.call}
    \mbox{call}(K, S, T) = e^{-q T}  S_0  N(d_1) - e^{-r T} K N(d_2) \,,
\end{equation}
\begin{equation}\label{eq:BSM.put}
    \mbox{put}(K, S, T) = e^{-r T} K N(-d_2) - e^{-q T} S_0 N(-d_1)  \, ,
\end{equation}
where $\sigma(K)$ are chosen to make the right hand side be equal to the known left hand side; $N$ and $d_{1,\, 2}$ are defined in Table \ref{tab:meaning.variable.callprice}. In other words, whenever the value of a $\mbox{call}(K)$ or a $\mbox{put}(K)$ is known, a unique $\sigma(K)$ makes equalities \eqref{eq:BSM.call} and \eqref{eq:BSM.put} hold for known $S_0$, $T$, $r$, $q$. The values of $\sigma(K)$ in formulae \eqref{eq:BSM.call} and \eqref{eq:BSM.put} are identical.
\par The ideal case when implied volatility is constant is equivalent to log-normality of the implied probability density of the future asset price $p(S_T)$,
$$\sigma(K) = \mbox{const}  \Leftrightarrow p(S_T) \, \mbox{is log-normal}\, .$$
It it straightforward to show that
\begin{equation}\label{eq:lemma.equilibrium}
    S_0 e^{-q T} n(d_1) = K e^{-r T} n(d_2) \, .
\end{equation}
\par The vanna-volga approach can be used to interpolate and extrapolate implied volatility given its values at three strikes \cite{Castagna.Mercurio:2007} in a parametric-free way.

\subsection{Stereographic projection}
\par
\begin{figure}[hbt!] % [h!]
\centering
\includegraphics[scale=0.35]{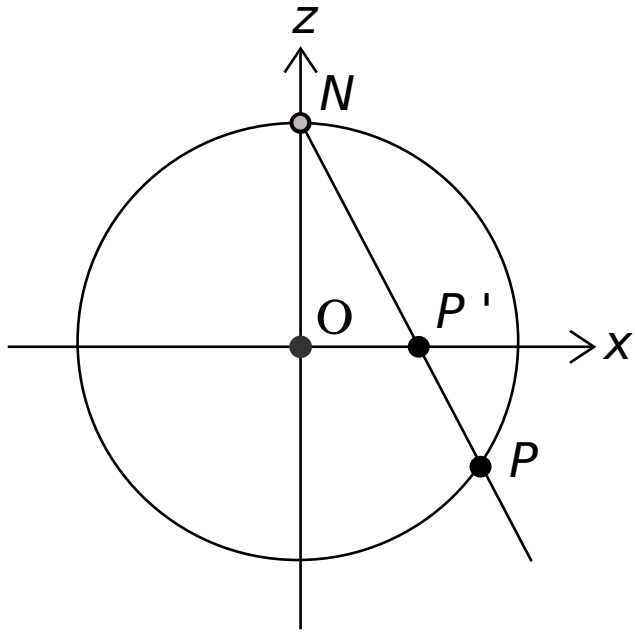} % , bb=0 0 85 48
\caption{A slice of stereographic projection that belongs to the plane $y = 0$. Point $P$ on the circle is projected into the point $P'$ on the $x$ axis.}
\label{Fig:demonstration.stereographic.projection}
\end{figure}

 Stereographic projection is a one-to-one mapping of a 3D sphere without its north pole onto the plane $z = 0$. Figure \ref{Fig:demonstration.stereographic.projection} demonstrates how a slice of stereographic projection that belongs to the plane $y = 0$ maps a unit circle without its north pole onto the $x$ axis.

%
% Edit SVG file here:
% https://www.vecteezy.com/editor/random
% Convert svg to eps here
% https://image.online-convert.com/convert-to-eps
%

The formulae for the stereographic projection and its inverse are well known. For the case of the slice in the plain $y = 0$ the transformation is applied to a unit circle centered at the origin $\{x^2 + z^2 = 1,\, y = 0\}$ and the relationships between the coordinates of the circle $(x,\, z)$ and projections $X$ on the $OX$ axis are as follows:

\begin{eqnarray}
    \nonumber
  X &=& \frac{x}{1 - z} \\
  \left(x,\, z\right) &=& \left(\frac{2 X}{1 + X^2},\, \frac{-1 + X^2}{1 + X^2} \right)\, .
  \label{eq:Cartesian.from.X}
\end{eqnarray}

\subsection{Euclidean and similarity curvatures}
\par Circles have constant Euclidean curvature. Euclidean curvature of a curve is invariant under rotations and translations of a curve while similarity curvature is invariant under rotations, translations and uniform scaling. Closeness of Euclidean curvature of a curve to constancy characterizes curve's ``closeness'' to being circular. Analysis of similarity curvature allows scale-invariant characterization of curves.
\par For an arbitrarily parameterized curve $\{x(t),\, y(t)\}$ the formulae for the Euclidean and similarity curvatures respectively are as follows \cite{Shirokovy_1959}:
$$\kappa_{E} = \frac{\dot{x} \ddot{y} - \dot{y}\ddot{x}}{\left(\dot{x}^2 + \dot{y}^2\right)}\, , $$

$$\kappa_{s} = 3 \frac{\dot{x} \ddot{x} + \dot{y} \ddot{y}}{\dot{x} \ddot{y} - \dot{y}\ddot{x}} - \frac{\dot{x} \dddot{y} - \dot{y}\dddot{x}}{\left(\dot{x} \ddot{y} - \dot{y}\ddot{x}\right)^2} \left(\dot{x}^2 + \dot{y}^2 \right) \, . $$

\subsection{Divergence between distributions}
\par Kullback-Leibler divergence \cite{CoverT.ThomasJ:91} (KL)
\begin{equation}\nonumber
    \mbox{KL}(P || Q) = \int_{-\infty}^{\infty} p(x) \log \left(\frac{p(x)}{q(x)} \right) dx
\end{equation}
is used to estimate discrepancy between two distributions; for example, between the Gamma distribution and the distribution represented by a circle translated from the origin. In the current work integration is implemented numerically based on the trapezoid rule and computation is based on natural logarithm so that the divergence is measured in nats.

\section{Methods}
\par Given a market state and time to expiry, that is the values of $r$, $q$, $T$, $S_0$, a one-to-one correspondence can be established between the set of implied volatility profiles $\{\sigma(K)\}$ and the set of probability densities\footnote{Some implied volatility profiles correspond to densities that may take negative values. So density here means an integrable function with finite expectation and whose integral over $\mathbb{R}_{+}$ is equal to 1.} $\{p(K)\}$ supported on $\mathbb{R}_{+}$ and having finite expectation via the formulae \eqref{eq:callprice}, \eqref{eq:2nd.derivative.call}, and \eqref{eq:BSM.call}. Earlier work \cite{Polyakov:2021} provides explanation, derivations and examples that include popular distributions, like gamma and uniform. Here implied volatility profiles are used to establish geometric representations of corresponding probability distributions.
\subsection{Representation of implied volatility smile by a curve in polar coordinate system}

\par
\begin{figure}[hbt!] % [h!]
\centering
% The plots were implemented in the code demonstration_stereographic_projection.m .
\includegraphics[scale=0.75]{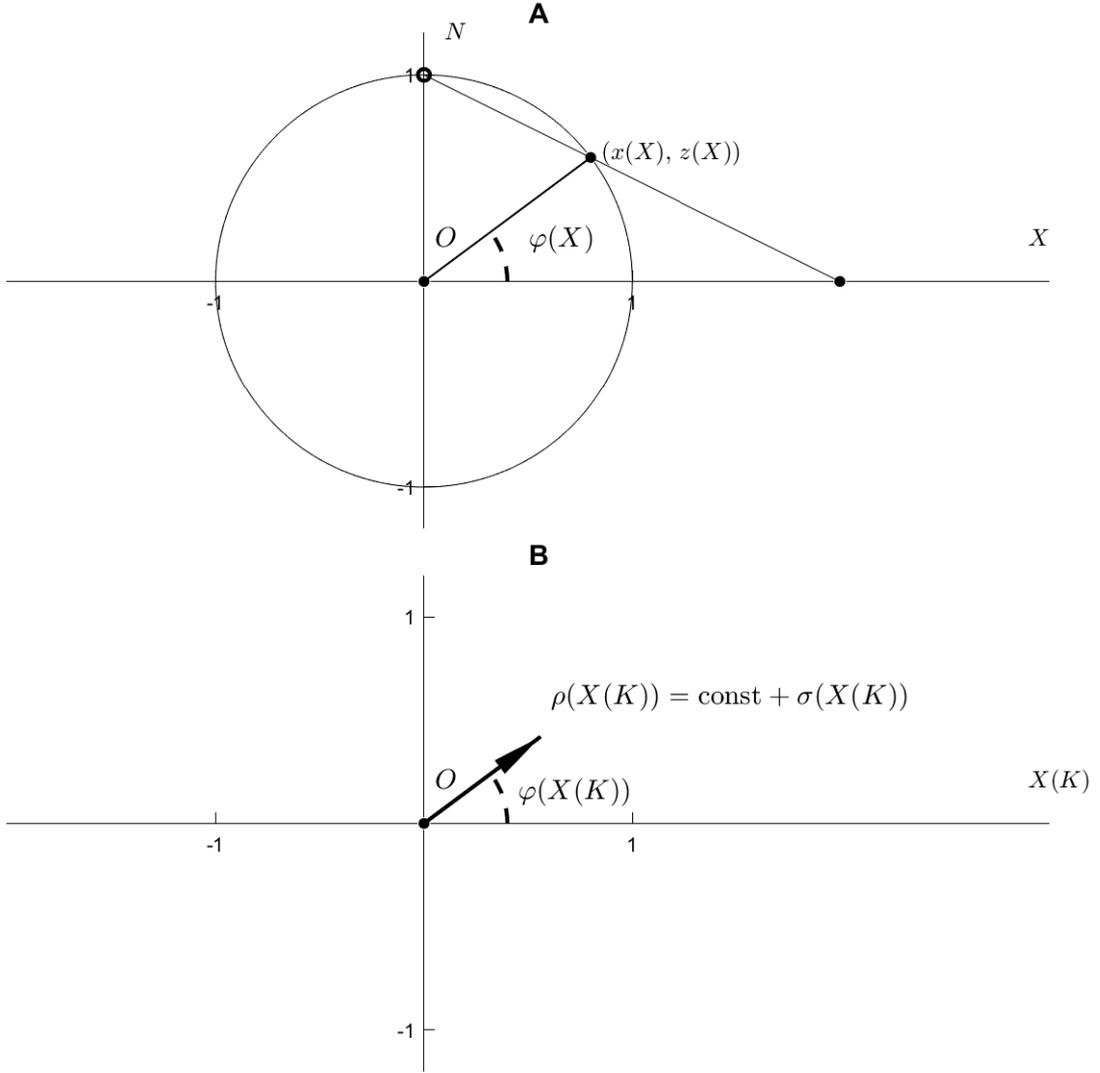}
\caption{Construction of geometric representation for implied volatility based on polar coordinate system. \textbf{A:} the strike $K$ is being transformed to corresponding polar angle $\varphi(X(K))$. \textbf{B:} implied volatility for that strike is being transformed to corresponding value of polar radius.}
\label{Fig:sigma2signature}
\end{figure}

 The proposed geometric representation of implied volatility employs the following steps:
\begin{enumerate}%${Lis:list.algorithm.stereographic}
  \item\label{label.R} For a given strike $K$ compute the value $X(K) = \frac{1}{R} \cdot \ln\frac{K}{\mbox{ATM}_{\mbox{rn}}}$. Here the constant $R$ controls how much of accumulated probability is supported for the strikes whose values $X(K)$ fall within the unit circle ($-1 < X(K) < 1$) using the stretching/extension of $\ln K$ by $\frac{1}{R}$; while $R$ is the only free parameter of the proposed methodology.
  \item Use equation \eqref{eq:Cartesian.from.X} to compute the Cartesian coordinates $(x,\, z)$ of the unique point on the unit circle corresponding to $X(K)$.
  \item Compute the polar angle $\varphi$ between the $OX$ axis and the line connecting the origin to the point $(x,\, z)$ on the circle as demonstrated in Figure \ref{Fig:sigma2signature}A.
  \item Compute the radial component $\rho(K) = \mbox{const} + \sigma(K)$  as demonstrated in Figure \ref{Fig:sigma2signature}B. Straightforward numeric computation leads to the function $\sigma(K)$ corresponding to some known probability distribution $p(K)$ \cite{Polyakov:2021}. I assume here that the values of $\sigma(K)$ are known with arbitrary good accuracy. At present the implementation sets $\mbox{const} = R$ from step \ref{label.R}.
  \item Geometrically represent the implied volatility profile $\sigma(K)$ with the curve $\{\rho(K) \cdot \cos \varphi, \, \rho(K) \cdot \sin \varphi\}$; this curve also represents the probability distribution $p(K)$ associated with $\sigma(K)$.
\end{enumerate}
In such a way the profile of implied volatility $\sigma(K)$ is mapped into a curve in the plane $xz$. For the presented construction, log-normal probability distribution whose implied volatility profile $\sigma(K)$ is constant is mapped into a curve with constant $\rho(K)$, in other words into a circle centered at the origin. For non log-normal distribution $\rho(K)$ is not constant and so its geometric representation does not correspond to a circle centered at the origin.

\par The procedure for geometric representation of probability distribution was applied to several gamma distributions with bell-shaped density profiles  and uniform distributions that do not have bell-shaped density profile, both supported on $\mathbb{R}_{+}$. Normal distribution and translated Student's $t$-distribution have bell-shaped probability density profile and can be picked in a way that the amount of the part supported outside of $\mathbb{R}_{+}$, that is $P(0)$, is as negligible as necessary by choosing sufficiently large positive expectation and sufficiently small standard deviation. Therefore, geometric representations of normal and translated Student's $t$ distributions with small enough $P(0)$ were also analyzed for the part supported on $\mathbb{R}_{+}$. Values of call options associated with  probability densities of the considered distributions are needed to find $\sigma(K)$ corresponding to $p(K)$ and vice versa. The formulae for call options are summarized in Table \ref{tab:formulae.expectation.callprice}.
\par Market provides options' prices for a small set of strikes and so the values of implied volatility are known at those points as well. Fitting curves to corresponding points in the representation space allows to implement continuous completion of implied volatility and to construct implied distribution.

\begin{table}[h]%[hbt!]
\caption{Formulae for call prices corresponding to a number of probability density functions.}
\begin{tabular}{|c|c|c|c|c|}
  \hline \hline
  Distribution  & Density function & Expected value & $\mbox{ATM}_{\mbox{rn}}$ & Call price \\
  & & = ATMF &  &  \\
  & & $= S_0 \cdot e^{(r - q) T}$ & & \\
  \hline  \hline
    Log-normal          &   $\frac{1}{x s \sqrt{2 \pi}}e^{-\frac{(\ln x - \mu)^2}{2 s^2}}$ &  $e^{\mu + \frac{s^2}{2}}$, & $e^{\mu + s^2}$ & Formula \eqref{eq:BSM.call} by Black-Scholes-Merton \\
   & & $\mu = \ln S_0 + $              & &    \\
   && $(r - q) T - \frac{s^2}{2}$ & & \\
   \hline
  Gamma              & $\frac{x^{\kappa-1} e^{-\frac{x}{\theta}}}{\theta^{\kappa} \Gamma(\kappa)}$ & $\kappa \cdot \theta$  & No closed & $e^{-r T} \left\{\theta \cdot \kappa \cdot \left[1 - P(K; \kappa + 1, \theta) \right] - \right.$ \\
  & & & form & $\left. K \cdot \left[1 - P(K; \kappa, \theta) \right] \right\}$ \; \cite{Polyakov:2021} \\
  & &  & formula & \\
  \hline
  Normal             &   $\frac{1}{s \sqrt{2 \pi}}e^{-\frac{( x - \mu)^2}{2 s^2}}$ &    $\mu$   & $\mu$        & $e^{-r T} \left[\left(\mbox{ATMF}(T) - K \right) \cdot N\left(\frac{\mbox{ATMF}(T) - K}{\sigma_N \sqrt{T}} \right) + \right.$  \\
  & & & & $ \left. \sigma_N \cdot \sqrt{T} \cdot  N'\left(\frac{\mbox{ATMF}(T) - K}{\sigma_N \sqrt{T}} \right) \right]$ \; \cite{Iwasawa:2001}  \\
  & & & & \\
  \hline
  Translated  & $\frac{\Gamma\left( \frac{\nu + 1}{2} \right)}{\sqrt{\nu \pi}\, \Gamma \left( \frac{\nu}{2} \right)} \times $ & $\mu$ & $\mu$     & $e^{-r T}  \frac{\nu}{\nu - 1} \cdot \frac{\Gamma\left( \frac{\nu + 1}{2} \right)}{\sqrt{\nu \pi}\, \Gamma \left( \frac{\nu}{2} \right)} \cdot \left[ 1 + \frac{(\mu - K)^2}{\nu} \right]^{\frac{1 - \nu}{2}} + $   \\
  Student & $\left( 1 + \frac{\left(x - \mu \right)^2}{\nu} \right) ^ {-\frac{\nu + 1}{2}}$ &&& $e^{-r T} \frac{\mu - K}{2} \cdot
     \begin{cases}
        I_{y(K)} \left( \frac{\nu}{2},\, \frac{1}{2}\right)    & K  \geq  \mu \\
        & \\
     2 -I_{y(K)} \left( \frac{\nu}{2},\, \frac{1}{2}\right)  &  K < \mu
  \end{cases}
$  \; \cite{Polyakov:2021} \\
  & & & & \\
  \hline
  &  &     & & \\
   Uniform & $\begin{cases}
   \frac{1}{b - a} &  a \leq x \leq  b  \\
   0       & \mbox{otherwise}
  \end{cases}$ &  $\frac{a + b}{2}$ & $a + \frac{b - a}{\sqrt{2}}$ & $ e^{-r T} \cdot \begin{cases}
    (b + a) / 2 - K & K  \leq a \\
     (b - K)^2 / \left[2 (b - a)\right] & a < K < b \\
    0 & b \leq K
    \end{cases}$  \; \cite{Polyakov:2021} \\
  & & & &  \\
  & & & &  \\
  & & & &  \\
  \hline
\end{tabular}
\label{tab:formulae.expectation.callprice}
\end{table}

\subsection{Fitting circle and ellipse to geometric representation}
\par Once geometric representation of a continuous distribution or of the point-wise volatility smile is known, it can be fit with a circle. Three points are used to fit a circle:
\begin{enumerate}
  \item $\{K_1,\, \sigma(K_1)\}: e^{q T} |\Delta_{\mbox{put}}(K_1)| = N(-d_1(K_1)) = 0.25$,
  \item $\{K_2,\, \sigma(K_2)\}: K_2 = \mbox{ATM}_{\mbox{rn}}$,
  \item $\{K_3,\, \sigma(K_3)\}: e^{q T} |\Delta_{\mbox{put}}(K_3)| = N(-d_1(K_3)) = 0.75$.
\end{enumerate}
To fit an ellipse, the following 5 points can be used:
\begin{enumerate}
  \item $\{K_1,\, \sigma(K_1)\}: e^{q T} |\Delta_{\mbox{put}}(K_1)| = N(-d_1(K_1)) = 0.1$,
  \item $\{K_1,\, \sigma(K_1)\}: e^{q T} |\Delta_{\mbox{put}}(K_1)| = N(-d_1(K_1)) = 0.25$,
  \item $\{K_2,\, \sigma(K_2)\}: K_2 = \mbox{ATM}_{\mbox{rn}}$,
  \item $\{K_3,\, \sigma(K_3)\}: e^{q T} |\Delta_{\mbox{put}}(K_3)| = N(-d_1(K_3)) = 0.75$,
  \item $\{K_3,\, \sigma(K_3)\}: e^{q T} |\Delta_{\mbox{put}}(K_3)| = N(-d_1(K_3)) = 0.9$.
\end{enumerate}
In case of market smile, the values of $K$ and $\sigma$ may correspond to $|\Delta_{\mbox{put}}(K)|$ without multiplication by $e^{q T}$.

\clearpage
\section{Results}
This section contains examples of applying the proposed methodology to probability distributions of different kinds and to a set of point-wise implied volatility data from the market. Two to three figures correspond to each considered distribution.
\begin{enumerate}
    \item Geometric representation of the considered distribution, circle and ellipse fitted to this geometric representation and 3/5 points used to fit circle/ellipse. The figure also includes geometric representation of the distribution defined by the implied volatility obtained with vanna-volga method applied to the three points used to fit the circle.
    \item Volatility smiles corresponding to the objects from the previous item, their corresponding probability density functions and difference between density of the considered distribution and distributions corresponding to the fitted circle and to vanna-volga based implied volatility profile. Kullbak-Leibler divergence for pairs of distributions is provided as well.
    \item Similarity and Euclidian curvatures of the geometric representation as a function of the polar angle with resect to the center of the fitted circle and as a function of $N(-d_{1})$. Computation of curvature is not applicable to the sparse point-wise input from the market and therefore only two figures characterize example from the market.
\end{enumerate}

%
% The figure was created using the function vols2sphere_different_distributions.m
% with gamma distribution.
% distributionS.gammA.What_fixed: 'sigma'
% count_std_normal = 9
% count_smile = 6
%
%distributionS =
%
%  struct with fields:
%
%                       expectation: 3.2768
%                          gammastd: 1.4482
%                 Type_distribution: 'Gamma'
%                             gammA: [1×1 struct]
%                        student_nu: 3.8229
%           How_to_accumulate_delta: 'differentiation_based_delta'
%                             theta: 0.6400
%                             kappa: 5.1200
%                   mixedlognormalS: [1×1 struct]
%                          variance: 2.0972
%                           Strikes: [1×5001 double]
%    SIgmabs_from_givendistribution: [6×5001 double]

\subsection{Examples of distributions supported on $\mathbb{R}_{+}$}

\par
\begin{figure}[hbt!]% [h!]
\centering
\includegraphics[scale=0.9]{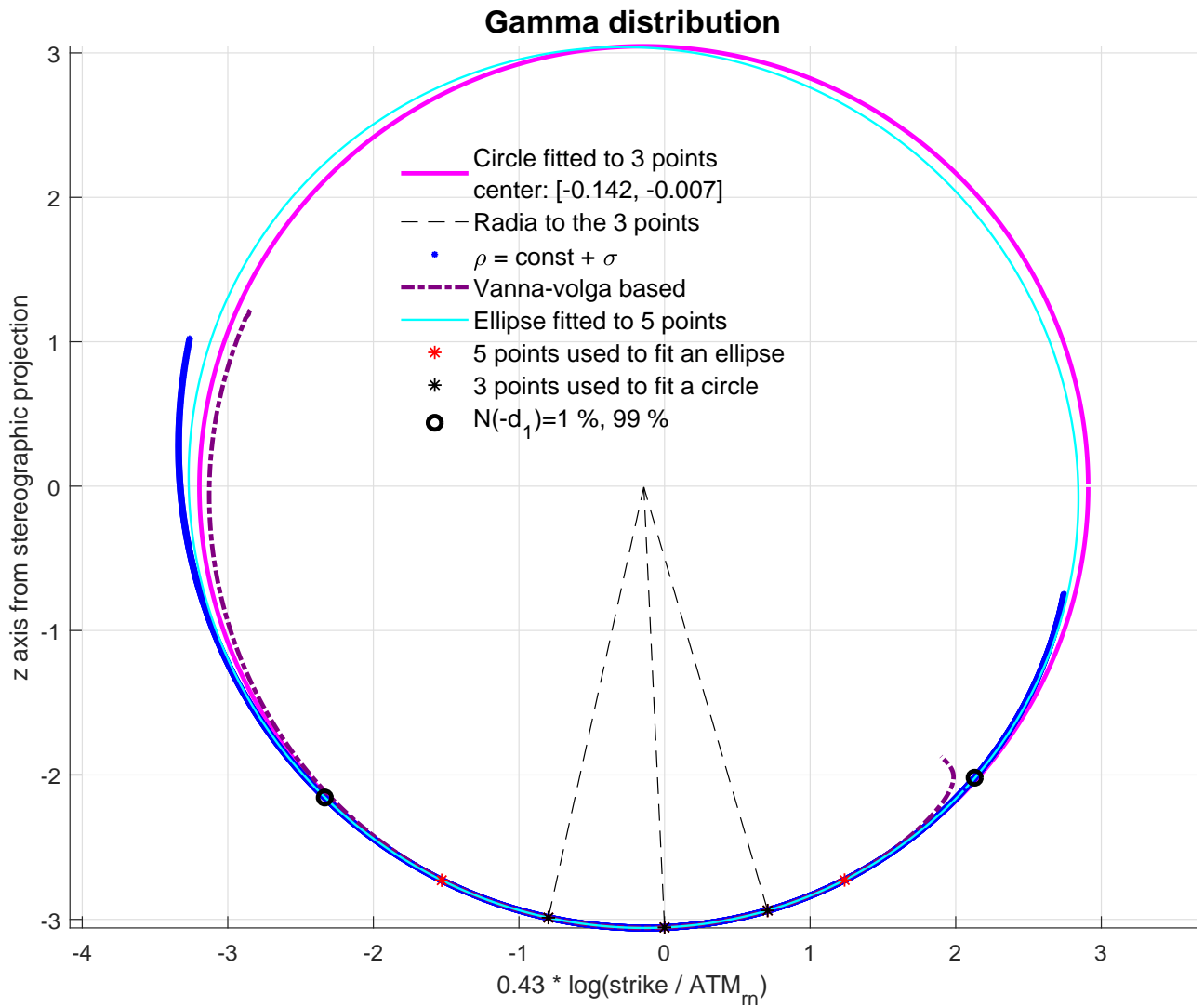}
\caption{Illustration of how a distribution characterized with its implied volatility profile can be approximated by inverting the circular representation. Geometric representation of gamma distribution with $\kappa = 5.12,\, \theta = 0.64$ (blue asterisks) is approximated with a circle translated from the origin (magenta). The circle, in turn, defines another probability distribution that provides a close approximation to represented gamma distribution as demonstrated in Figure \ref{fig:example.pdf.gamma.smile}.}
\label{fig:example.circle.smile.gamma}
\end{figure}

\begin{figure}[hbt!] % [h!]
\centering
\includegraphics[scale=0.7]{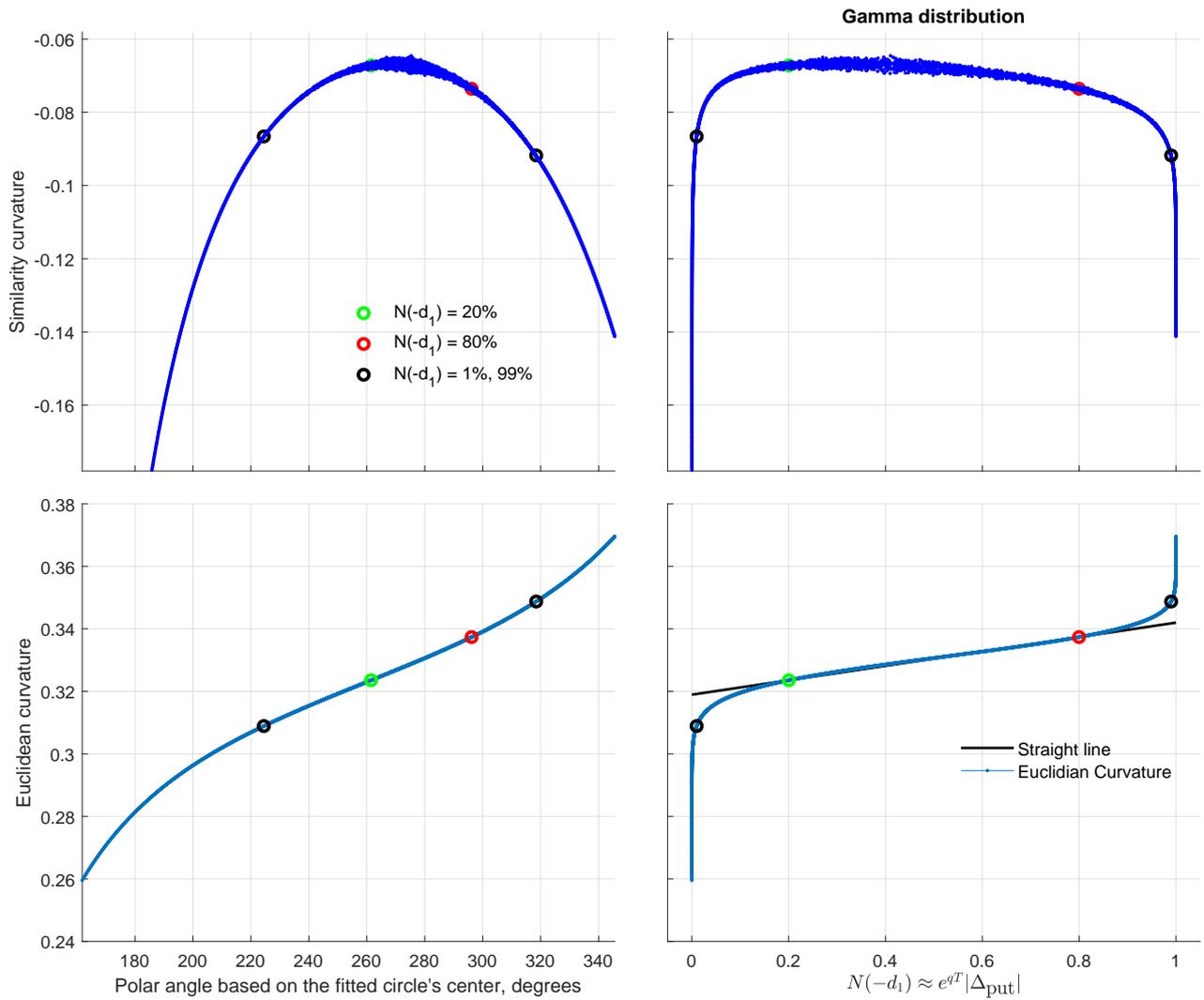}
\caption{Similarity and Euclidean curvatures of the geometric representation from Figure \ref{fig:example.circle.smile.gamma}.}
\label{fig:example.pdf.gamma.geomsignature}
\end{figure}

\begin{figure}[hbt!]% [h!]
\centering
\includegraphics[scale=0.85]{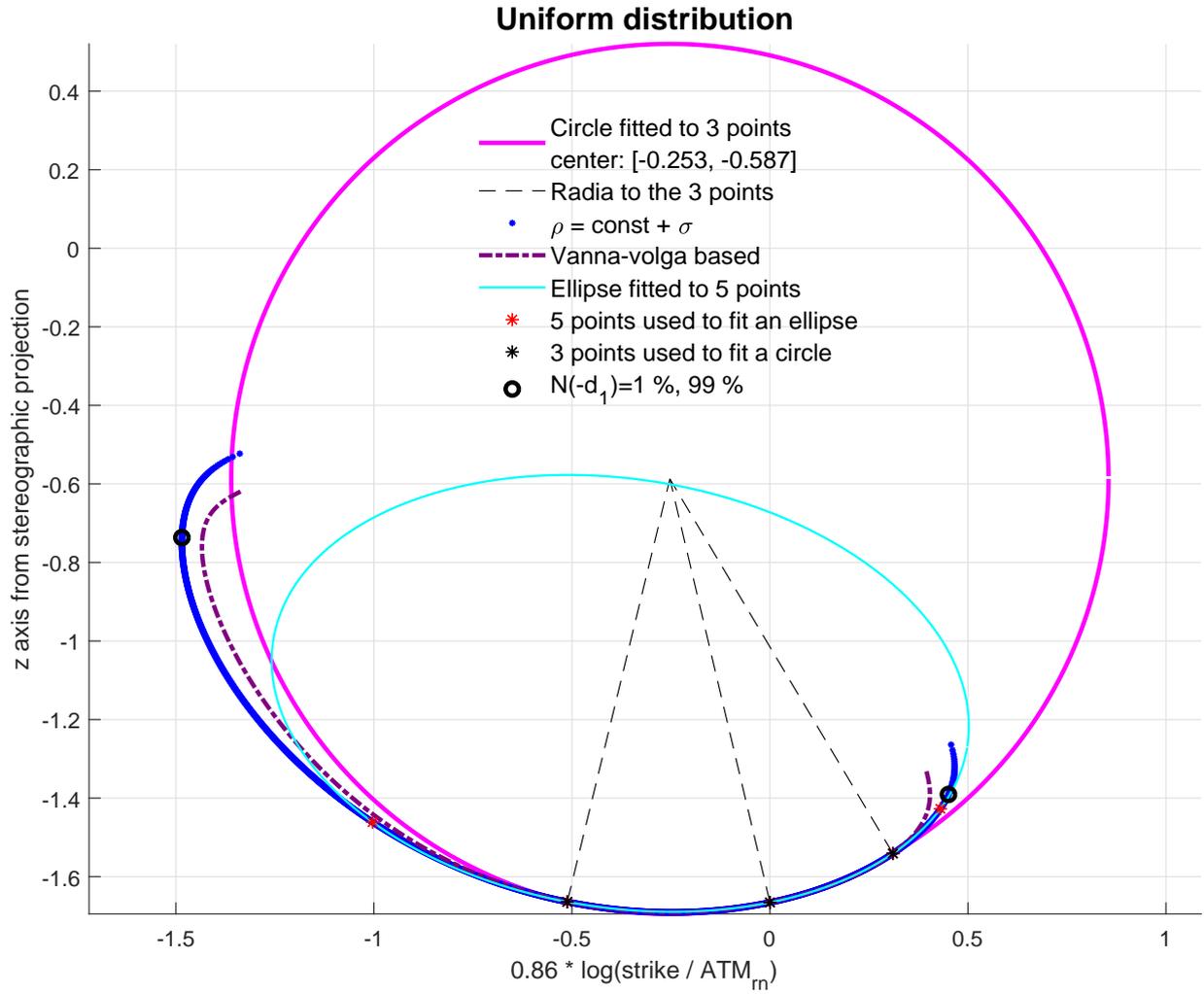}
\caption{Geometric representation of the uniform distribution $p(x) = 0.2887$ when $2.0109 \leq x \leq 5.4750$ and $p(x) = 0$ when $x \notin [2.0109,\, 5.4750]$ (blue asterisks) and corresponding circular representation that passes through the geometric representation at three points identified with $e^{q T} |\Delta_{put}| = 0.25, 0.5, 0.75$ (magenta). Probability density profile of a uniform distribution is not bell shaped, so it is not surprising that circular representation and its inverse do not provide a good approximation to uniform distribution.}
\label{fig:example.stereographic.uniform}
\end{figure}

\begin{figure}[hbt!] % [h!]
\centering
\includegraphics[scale=0.9]{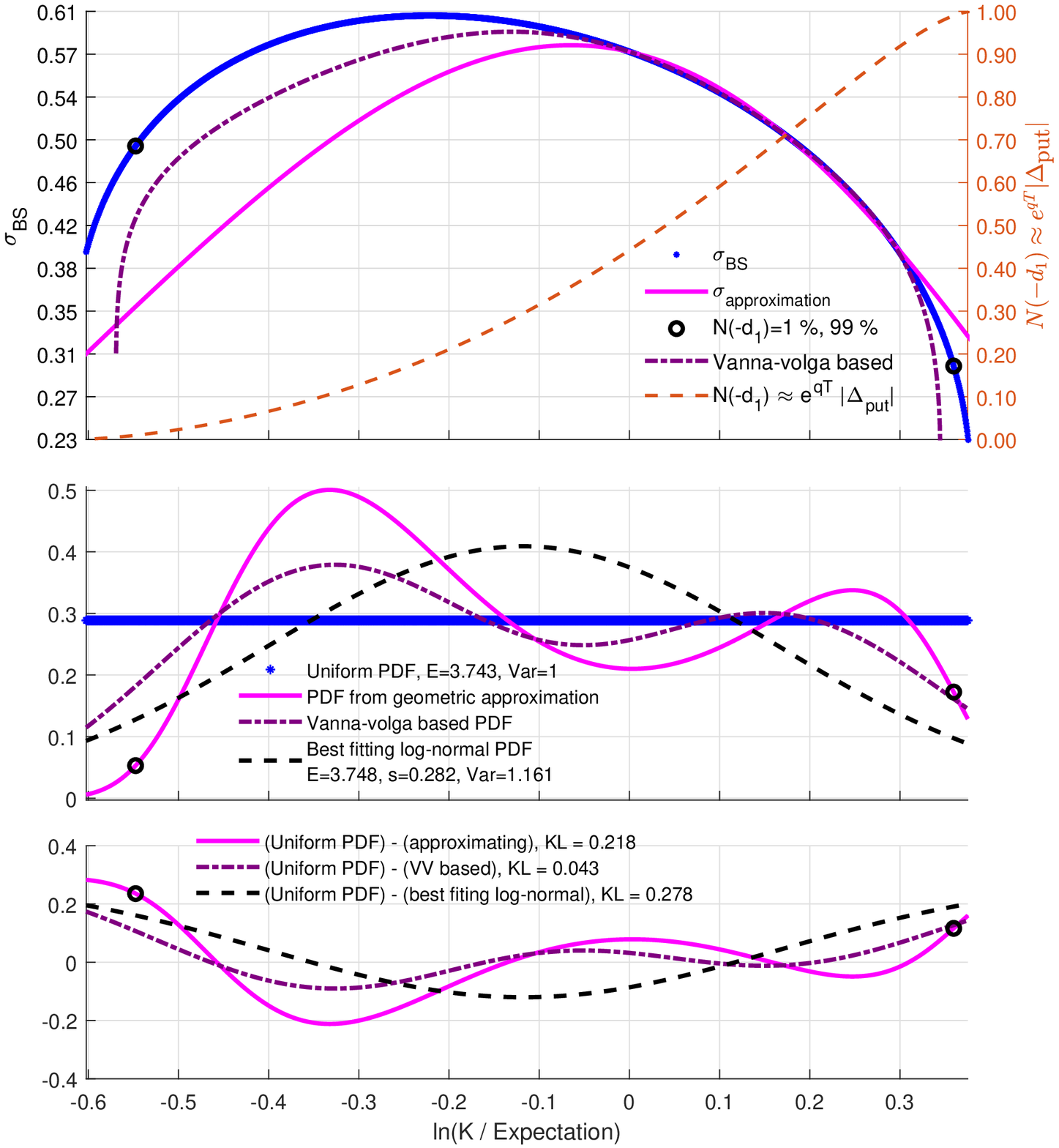}
\caption{Geometric representation of the uniform distribution from Figure \ref{fig:example.stereographic.uniform} and corresponding circular representation are further analyzed. \textbf{Upper part}: Volatility smile for the analyzed uniform distribution (blue), volatility smile derived from the circle fitted to the geometric representation (magenta) and implied volatility computed with the vanna-volga method (dashed violet). The two small black circles correspond to the strikes at which $N(-d_1) = 0.01,\, 0.99$.  \textbf{Middle part}: uniform PDF (blue) and PDF whose geometric representation is circular approximation (magenta) noticeably deviate; this is not surprising as uniform distribution does not have bell-shaped density profile. Log-normal PDF that provides the best fit to analyzed uniform PDF in the Kullback-Leibler sense is depicted with dashed blue.  The values of Kullback-Leibler divergence between the distributions are presented in the legend. \textbf{Below part}: Difference between analyzed uniform PDF and (1) PDF represented with the circle (magenta), (2) vanna-volga based PDF (dashed violet) and (3) best fitting log-normal PDF (dashed black).}
\label{fig:example.pdf.uniform.smile}
\end{figure}

\begin{figure}[hbt!] % [h!]
\centering
\includegraphics[scale=0.7]{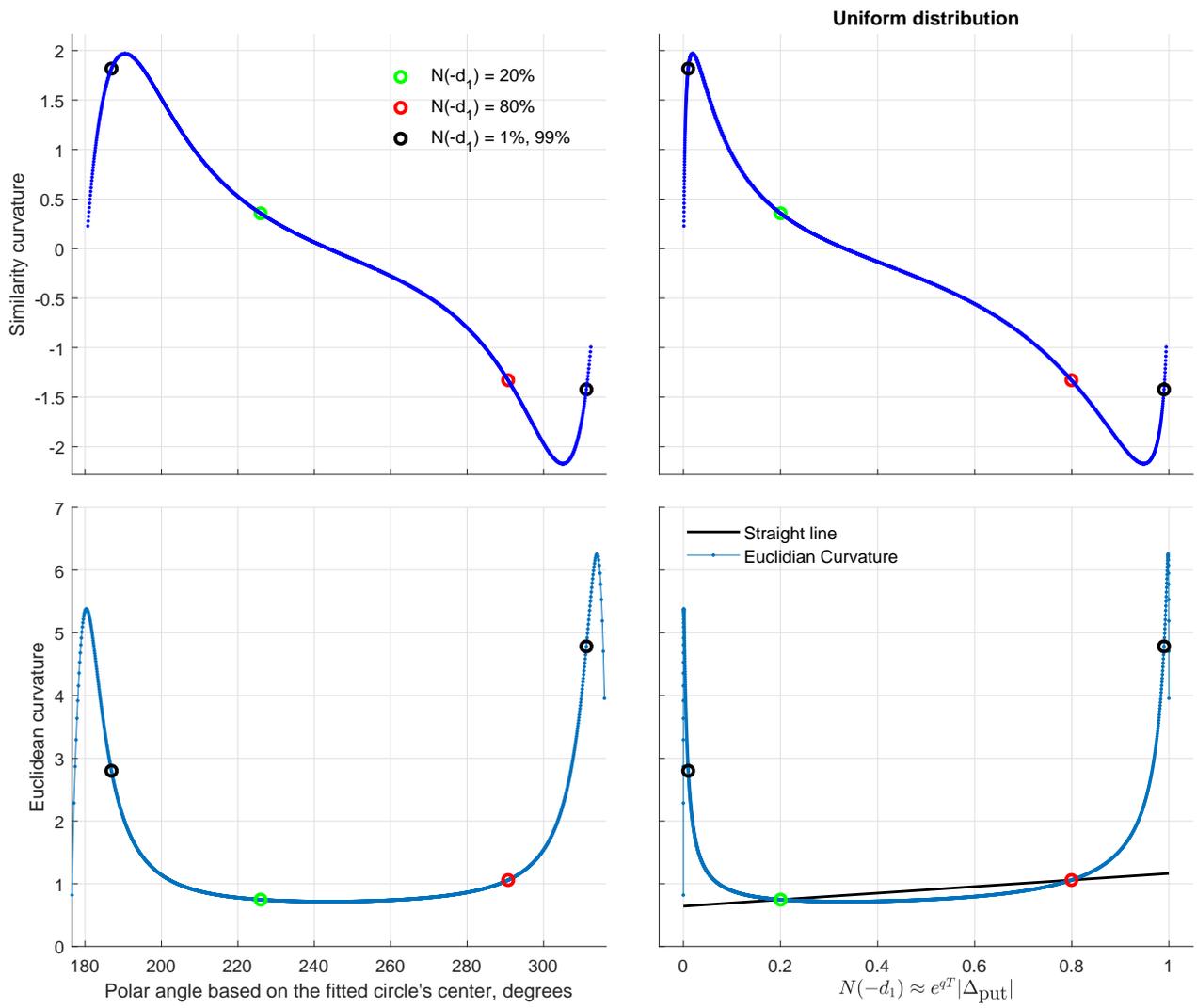}
\caption{Similarity and Euclidean curvatures of the geometric representation of the uniform distribution from Figure \ref{fig:example.stereographic.uniform}.}
\label{fig:example.pdf.uniform.geomsignature}
\end{figure}

 Gamma distributions and specifically chosen uniform distributions are supported on $\mathbb{R}_{+}$. While rich subset of gamma distributions consists of densities with bell-shaped profile, the profile of uniform distributions is non bell-shaped. Typically Gamma distributions with bell-shaped density profile are well approximated with distributions represented by fitted circles as demonstrated in Figures \ref{fig:example.circle.smile.gamma}-\ref{fig:example.pdf.gamma.geomsignature}; however uniform distributions, as expected, cannot be well approximated with distributions represented by fitted circles as demonstrated in Figures \ref{fig:example.stereographic.uniform}-\ref{fig:example.pdf.uniform.geomsignature}.

The divergence between the Gamma and circle based approximating PDFs is much smaller than other divergences measured as can be seen in the legend to the below plot in Figure \ref{fig:example.pdf.gamma.smile}.

% The same distribution, fit to PDF.
\begin{figure}[hbt!] % [h!]
\centering
\includegraphics[scale=0.9]{Gamma_smile_KL_VV_example_countSTDnormal09_countSMILE06.eps}
\caption{Geometric representation of the gamma distribution from Figure \ref{fig:example.circle.smile.gamma} and its circular approximation are further analyzed. \textbf{Upper part}: Volatility smile for the analyzed gamma distribution (blue), volatility smile derived from the circle fitted to the geometric representation (magenta) and implied volatility computed with the vanna-volga method (dashed violet). The two small black circles correspond to the strikes at which $N(-d_1) = 0.01,\, 0.99$.  \textbf{Middle part}: gamma PDF (blue) and PDF whose geometric representation is circular approximation (magenta) can be hardly distinguished. Log-normal PDF that provides the best fit to analyzed gamma PDF in the Kullback-Leibler sense is depicted with dashed blue. The values of Kullback-Leibler divergence between the distributions are presented in the legend. \textbf{Below part}: Difference between the gamma PDF and (1) PDF represented with the circle (magenta), (2) vanna-volga based PDF (dashed violet) and (3) best fitting log-normal PDF (dashed black).}
\label{fig:example.pdf.gamma.smile}
\end{figure}

\clearpage% Flush earlier floats (otherwise order might not be correct)
%
% The figure was created using the function vols2sphere_different_distributions.m
% with uniform distribution.
% count_std_normal = 10
% count_smile = 5
%
%
%distribut%ionS =
%
%  struct with fields:
%
%                       expectation: 3.7430
%                          gammastd: 0.5055
%                 Type_distribution: 'Uniform'
%                             gammA: []
%                        student_nu: NaN
%           How_to_accumulate_delta: 'differentiation_based_delta'
%                             theta: 0.0683
%                             kappa: 54.8217
%                   mixedlognormalS: [1×1 struct]
%                          variance: 1
%                 uniform_b_minus_a: 3.4641
%                           Strikes: [1×5001 double]
%    SIgmabs_from_givendistribution: [5×5001 double]
%

% The same distribution, fit to PDF.

\clearpage% Flush earlier floats (otherwise order might not be correct)

\subsection{Geometric completion of implied volatility and partially observable probability, case of capital markets}

\par The values of implied volatility $\sigma(K,\, T)$ are cited in the market for a limited set of strikes $K$ and expiries $T$. Usually, such $K$ correspond to 3 points: (1) ATM and (2-3) $25\Delta$ call and put; sometimes additional values are provided at $10\Delta$ call/put, $35\Delta$ call/put, $5\Delta$ call/put. Known values of $\sigma$ are interpolated/extrapolated for arbitrary values of $K$ in order to price options at arbitrary strikes.
\par Three volatility surfaces are considered here, one for each of the three currency pairs: EURUSD, USDJPY and USDILS\footnote{Market data courtesy of Bloomberg LP.}. Volatility completion was implemented separately for each expiry based on circular representation and based on the vanna-volga approach; both methods used three volatility values, at (1) ATM and (2-3) $25\Delta$ call and put. The results of completion can be compared to the market values at $10\Delta$ call/put, $35\Delta$ call/put, $5\Delta$ call/put. As can be seen in Tables \ref{table:table.vol.discrepancies.circles.epxiryandtenor.EURUSD} - \ref{table:table.vol.discrepancies.vannavolga.epxiryandtenor.USDILS}, the discrepancies between the modeled and ``market'' volatilities are smaller for completion based on circular representation than the one based on the vanna-volga method.

% The figure was created using the function vols2sphere_how_to_run.m, EURUSD
%
\begin{figure}[hbt!] %[h!]
\centering
\includegraphics[scale=0.85]{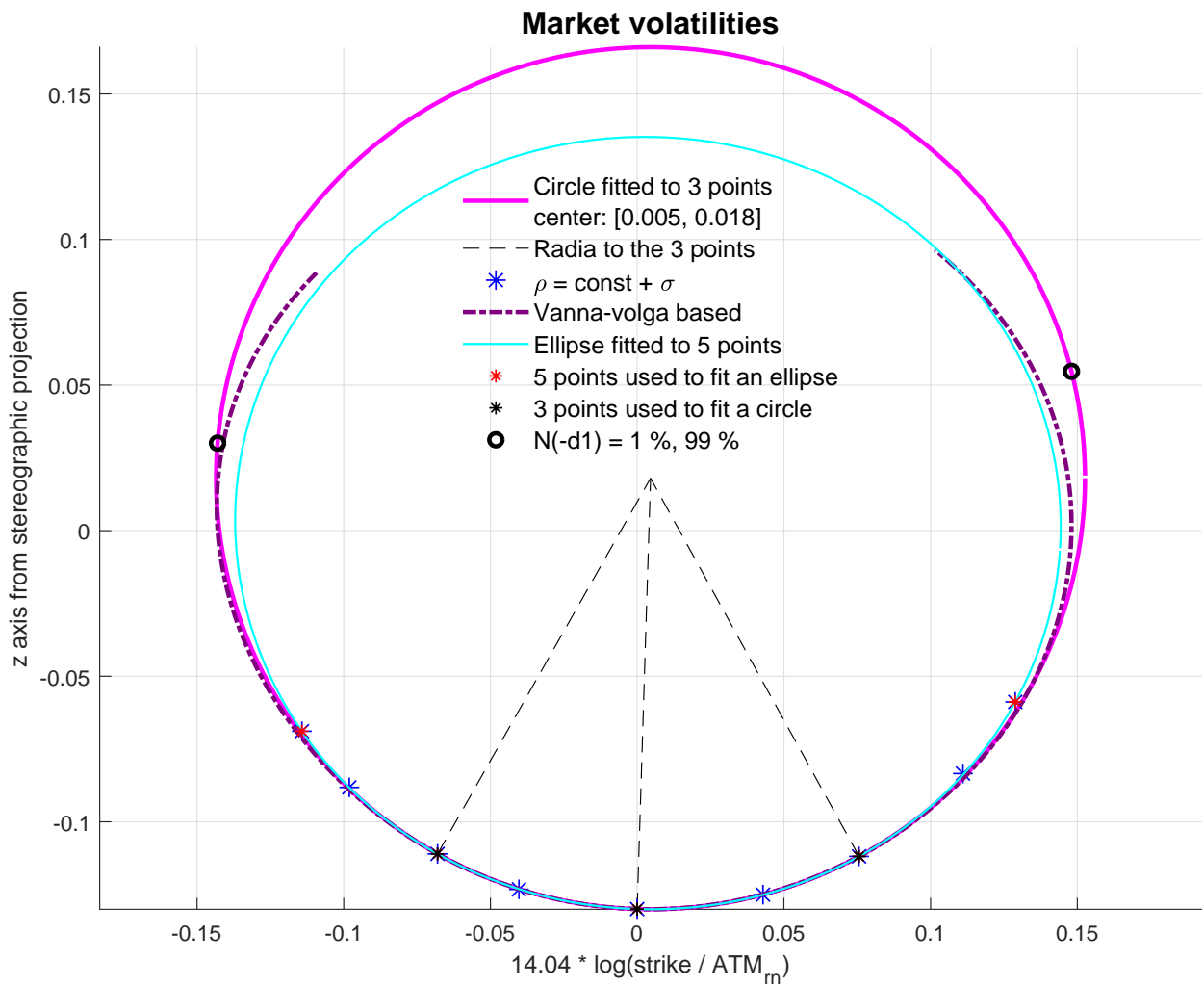} % fit_circle_3M.eps}
\caption{Illustration of fitting market smile with a circle: three points at the strikes corresponding to 25, 75 $\left|\Delta \mbox{put} \right|$ and ATM from the market smile are fit with a circle, the rest points of the smile are relatively close to the fitted circle. The values of the market smile are available for the strikes at 5, 10, 25, 35, 65, 75, 90, 95 $\left|\Delta \mbox{put}\right|$ and ATM. The market data are courtesy to Bloomberg LP.}
\label{fig:example.circle.smile}
\end{figure}

% The same distribution, fit to PDF.
\begin{figure}[hbt!] % [h!]
\centering
\includegraphics[scale=0.9]{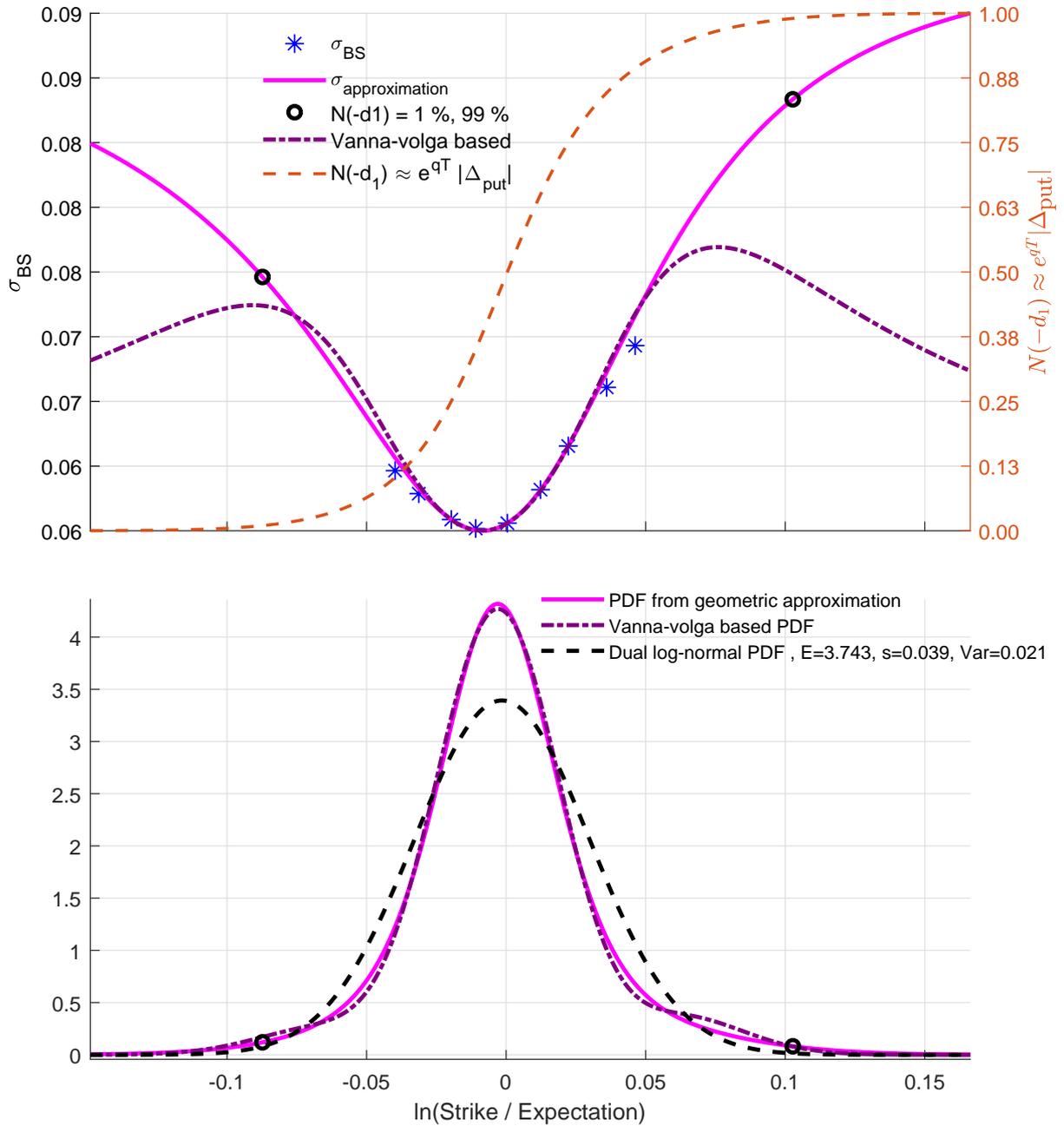}
\caption{Probability density function for the circular representation of the market data from Figure \ref{fig:example.circle.smile}. \textbf{Upper part}: Market smile (blue), smile implied from its geometric approximation with a circle (magenta) and Vanna-Volga based implied volatility (dashed dark purple). \textbf{Below part}: PDF resulting from circular representation (magenta), from Vanna-Volga based implied volatility (dashed dark purple) and dual log-normal PDF (dashed black).}
\label{fig:example.pdf.market.smile}
\end{figure}

\clearpage

\begin{table}
\caption{Discrepancies between the proposed volatilities and market volatilities, EURUSD; 0.0001 corresponds to 0.01\% of implied volatility.}
\addtolength{\tabcolsep}{-2pt}
\begin{tabular}{|c	|l	|l	|l	|l	|l	|l	|l	|l	|l	|l	|l|}
  \hline
Expiry&	$10 |\Delta_{\mbox{put}}|$ & 	$15 |\Delta_{\mbox{put}}|$ & 	$25 |\Delta_{\mbox{put}}|$ & 	$35 |\Delta_{\mbox{put}}|$ & 	ATM & 	$35 \Delta_{\mbox{call}}$ & 	$25 \Delta_{\mbox{call}}$ & 	$15 \Delta_{\mbox{call}}$ & 	$10 \Delta_{\mbox{call}}$ & 	 & 	$L^2$ norm \\ \hline
2W&	-0.0008&	-0.0004&	0&	0&	0&	0.0001&	0&	-0.0002&	-0.0005&	&	0.0011\\
3W&	-0.0007&	-0.0004&	0&	0&	0&	0.0001&	0&	-0.0003&	-0.0007&	&	0.0012\\
1M&	-0.0008&	-0.0005&	0&	0&	0&	0.0001&	0&	-0.0004&	-0.0008&	&	0.0013\\
2M&	-0.0011&	-0.0007&	0&	0&	0&	0.0001&	0&	-0.0002&	-0.0005&	&	0.0014\\
3M&	-0.0006&	-0.0005&	0&	-0.0001&	0&	0.0001&	0&	0.0001&	0.0001&	&	0.0008\\
4M&	-0.0002&	-0.0003&	0&	-0.0001&	0&	0.0001&	0&	0.0001&	0.0001&	&	0.0004\\
6M&	0&	-0.0002&	0&	-0.0001&	0&	0.0001&	0&	0.0004&	0.0005&	&	0.0007\\
9M&	0.0003&	-0.0001&	0&	-0.0002&	0&	0&	0&	0.0004&	0.0005&	&	0.0007\\
1Y&	0.0003&	-0.0001&	0&	-0.0001&	0&	0.0001&	0&	0.0007&	0.0009&	&	0.0012\\
18M&	0.0007&	0.0001&	0&	-0.0002&	0&	0.0001&	0&	0.0005&	0.0007&	&	0.0011\\
2Y&	0.0007&	0.0002&	0&	-0.0002&	0&	0.0001&	0&	0.0005&	0.0005&	&	0.0011\\
3Y&	0.0009&	0.0003&	0&	-0.0001&	0&	0.0001&	0&	0.0003&	0.0003&	&	0.0010\\
4Y&	0.0010&	0.0004&	0&	-0.0001&	0&	0&	0&	0.0003&	0.0003&	&	0.0012\\
5Y&	0.0011&	0.0005&	0&	-0.0001&	0&	0&	0&	0.0003&	0.0004&	&	0.0013\\
&	&	&	&	&	&	&	&	&	&	&	\\ \hline
$L^2$ norm&	0.0027&	0.0014&	0&	0.0004&	0&	0.0003&	0&	0.0014&	0.0020&	&	0.0040\\
\hline
\end{tabular}
\label{table:table.vol.discrepancies.circles.epxiryandtenor.EURUSD}
\end{table}

\begin{table}
\caption{Discrepancies between the vanna-volga volatilities and market volatilities, EURUSD; 0.0001 corresponds to 0.01\% of implied volatility.}
\addtolength{\tabcolsep}{-2pt}
\begin{tabular}{|c	|l	|l	|l	|l	|l	|l	|l	|l	|l	|l	|l|}
  \hline
Expiry&	$10 |\Delta_{\mbox{put}}|$ & 	$15 |\Delta_{\mbox{put}}|$ & 	$25 |\Delta_{\mbox{put}}|$ & 	$35 |\Delta_{\mbox{put}}|$ & 	ATM & 	$35 \Delta_{\mbox{call}}$ & 	$25 \Delta_{\mbox{call}}$ & 	$15 \Delta_{\mbox{call}}$ & 	$10 \Delta_{\mbox{call}}$ & 	 & 	$L^2$ norm \\ \hline
2W&	-0.0014&	-0.0008&	0&	0.0001&	0&	0.0001&	0&	-0.0004&	-0.0010&	&	0.0019\\
3W&	-0.0014&	-0.0008&	0&	0.0001&	0&	0&	0&	-0.0004&	-0.0011&	&	0.0020\\
1M&	-0.0015&	-0.0009&	0&	0.0001&	0&	0&	0&	-0.0004&	-0.0011&	&	0.0021\\
2M&	-0.0019&	-0.0012&	0&	0.0001&	0&	0.0001&	0&	-0.0002&	-0.0007&	&	0.0023\\
3M&	-0.0013&	-0.0011&	0&	0.0001&	0&	0&	0&	0.0003&	0.0003&	&	0.0017\\
4M&	-0.0008&	-0.0009&	0&	0.0001&	0&	0&	0&	0.0004&	0.0003&	&	0.0013\\
6M&	-0.0005&	-0.0008&	0&	0.0001&	0&	0&	0&	0.0007&	0.0009&	&	0.0014\\
9M&	0&	-0.0007&	0&	0&	0&	-0.0001&	0&	0.0008&	0.0010&	&	0.0015\\
1Y&	0&	-0.0008&	0&	0.0001&	0&	-0.0001&	0&	0.0012&	0.0016&	&	0.0022\\
18M&	-0.0001&	-0.0006&	0&	0&	0&	0&	0&	0.0008&	0.0009&	&	0.0013\\
2Y&	-0.0003&	-0.0007&	0&	0&	0&	0&	0&	0.0006&	0.0004&	&	0.0010\\
3Y&	-0.0007&	-0.0005&	0&	0&	0&	0&	0&	0&	-0.0006&	&	0.0011\\
4Y&	-0.0006&	-0.0003&	0&	0&	0&	0&	0&	-0.0002&	-0.0008&	&	0.0011\\
5Y&	-0.0005&	-0.0002&	0&	0&	0&	0.0001&	0&	-0.0002&	-0.0009&	&	0.0010\\
&	&	&	&	&	&	&	&	&	&	&	\\ \hline
$L^2$ norm&	0.0037&	0.0029&	0&	0.0003&	0&	0.0002&	0&	0.0020&	0.0034&	&	0.0061\\
  \hline
\end{tabular}
\label{table:table.vol.discrepancies.vannavolga.epxiryandtenor.EURUSD}
\end{table}

\begin{table}
\caption{Discrepancies between the proposed volatilities and market volatilities, USDJPY; 0.0001 corresponds to 0.01\% of implied volatility.}
\addtolength{\tabcolsep}{-2pt}
\begin{tabular}{|c	|l	|l	|l	|l	|l	|l	|l	|l	|l	|l	|l|}
  \hline
Expiry&	$10 |\Delta_{\mbox{put}}|$ & 	$15 |\Delta_{\mbox{put}}|$ & 	$25 |\Delta_{\mbox{put}}|$ & 	$35 |\Delta_{\mbox{put}}|$ & 	ATM & 	$35 \Delta_{\mbox{call}}$ & 	$25 \Delta_{\mbox{call}}$ & 	$15 \Delta_{\mbox{call}}$ & 	$10 \Delta_{\mbox{call}}$ & 	 & 	$L^2$ norm \\ \hline
2W&	-0.0021 & 	-0.0014 & 	0 & 	0.0002 & 	0 & 	0.0003 & 	0 & 	-0.0013 & 	-0.0024 & 	 & 	0.0038 \\
3W&	-0.0012 & 	-0.0009 & 	0 & 	0.0002 & 	0 & 	0.0003 & 	0 & 	-0.0011 & 	-0.0021 & 	 & 	0.0028 \\
1M&	-0.0007 & 	-0.0005 & 	0 & 	0 & 	0 & 	0.0002 & 	0 & 	-0.001 & 	-0.0019 & 	 & 	0.0023 \\
2M&	0.0003 & 	-0.0001 & 	0 & 	-0.0001 & 	0 & 	0.0001 & 	0 & 	-0.0008 & 	-0.0016 & 	 & 	0.0018 \\
3M&	-0.0003 & 	-0.0004 & 	0 & 	-0.0001 & 	0 & 	0.0001 & 	0 & 	-0.0008 & 	-0.0016 & 	 & 	0.0019 \\
4M&	0 & 	-0.0003 & 	0 & 	-0.0001 & 	0 & 	0 & 	0 & 	-0.0007 & 	-0.0015 & 	 & 	0.0017 \\
6M&	-0.0004 & 	-0.0005 & 	0 & 	-0.0001 & 	0 & 	0 & 	0 & 	-0.0008 & 	-0.0017 & 	 & 	0.002 \\
9M&	0.0002 & 	-0.0002 & 	0 & 	-0.0001 & 	0 & 	-0.0001 & 	0 & 	-0.0005 & 	-0.001 & 	 & 	0.0012 \\
1Y&	-0.0004 & 	-0.0005 & 	0 & 	-0.0001 & 	0 & 	-0.0001 & 	0 & 	-0.0005 & 	-0.0011 & 	 & 	0.0013 \\
18M&	0.0006 & 	0 & 	0 & 	-0.0001 & 	0 & 	-0.0002 & 	0 & 	0.0001 & 	-0.0002 & 	 & 	0.0006 \\
2Y&	0.0005 & 	0 & 	0 & 	-0.0002 & 	0 & 	-0.0004 & 	0 & 	0.0004 & 	0.0004 & 	 & 	0.0009 \\
3Y&	0.0004 & 	-0.0001 & 	0 & 	-0.0003 & 	0 & 	-0.0005 & 	0 & 	0.0005 & 	0.0005 & 	 & 	0.001 \\
4Y&	0.0009 & 	0.0002 & 	0 & 	-0.0003 & 	0 & 	-0.0007 & 	0 & 	0.0008 & 	0.0009 & 	 & 	0.0017 \\
5Y&	0.0013 & 	0.0004 & 	0 & 	-0.0004 & 	0 & 	-0.0009 & 	0 & 	0.0009 & 	0.0011 & 	 & 	0.0022 \\
&	 & 	 & 	 & 	 & 	 & 	 & 	 & 	 & 	 & 	 & 	 \\  \hline
$L^2$ norm&	0.0032 & 	0.002 & 	0 & 	0.0007 & 	0 & 	0.0014 & 	0 & 	0.003 & 	0.0053 & 	 & 	0.0074 \\
  \hline
\end{tabular}
\label{table:table.vol.discrepancies.circles.epxiryandtenor.USDJPY}
\end{table}

\begin{table}
\caption{Discrepancies between the vanna-volga volatilities and market volatilities, USDJPY; 0.0001 corresponds to 0.01\% of implied volatility.}
\addtolength{\tabcolsep}{-2pt}
\begin{tabular}{|c	|l	|l	|l	|l	|l	|l	|l	|l	|l	|l	|l|}
  \hline
Expiry&	$10 |\Delta_{\mbox{put}}|$ & 	$15 |\Delta_{\mbox{put}}|$ & 	$25 |\Delta_{\mbox{put}}|$ & 	$35 |\Delta_{\mbox{put}}|$ & 	ATM & 	$35 \Delta_{\mbox{call}}$ & 	$25 \Delta_{\mbox{call}}$ & 	$15 \Delta_{\mbox{call}}$ & 	$10 \Delta_{\mbox{call}}$ & 	 & 	$L^2$ norm \\ \hline
2W&	-0.0033 & 	-0.0021 & 	0 & 	0.0003 & 	0 & 	0.0002 & 	0 & 	-0.0015 & 	-0.003 & 	 & 	0.0052 \\
3W&	-0.0027 & 	-0.0017 & 	0 & 	0.0003 & 	0 & 	0.0002 & 	0 & 	-0.0012 & 	-0.0025 & 	 & 	0.0043 \\
1M&	-0.0024 & 	-0.0015 & 	0 & 	0.0002 & 	0 & 	0.0001 & 	0 & 	-0.0011 & 	-0.0023 & 	 & 	0.0038 \\
2M&	-0.0014 & 	-0.0012 & 	0 & 	0.0002 & 	0 & 	0 & 	0 & 	-0.0006 & 	-0.0016 & 	 & 	0.0025 \\
3M&	-0.0019 & 	-0.0015 & 	0 & 	0.0002 & 	0 & 	-0.0001 & 	0 & 	-0.0009 & 	-0.0021 & 	 & 	0.0033 \\
4M&	-0.0015 & 	-0.0014 & 	0 & 	0.0002 & 	0 & 	-0.0001 & 	0 & 	-0.0008 & 	-0.002 & 	 & 	0.003 \\
6M&	-0.0016 & 	-0.0016 & 	0 & 	0.0002 & 	0 & 	-0.0001 & 	0 & 	-0.0009 & 	-0.0024 & 	 & 	0.0034 \\
9M&	-0.0008 & 	-0.0013 & 	0 & 	0.0001 & 	0 & 	-0.0002 & 	0 & 	-0.0006 & 	-0.0019 & 	 & 	0.0025 \\
1Y&	-0.0006 & 	-0.0014 & 	0 & 	0.0002 & 	0 & 	-0.0002 & 	0 & 	-0.0008 & 	-0.0023 & 	 & 	0.0029 \\
18M&	0.0005 & 	-0.0009 & 	0 & 	0.0001 & 	0 & 	-0.0003 & 	0 & 	-0.0004 & 	-0.0017 & 	 & 	0.0021 \\
2Y&	0.0007 & 	-0.001 & 	0 & 	0 & 	0 & 	-0.0004 & 	0 & 	-0.0002 & 	-0.0013 & 	 & 	0.0018 \\
3Y&	0.001 & 	-0.001 & 	0 & 	0 & 	0 & 	-0.0005 & 	0 & 	-0.0002 & 	-0.0013 & 	 & 	0.002 \\
4Y&	0.0014 & 	-0.0008 & 	0 & 	0 & 	0 & 	-0.0007 & 	0 & 	0 & 	-0.0009 & 	 & 	0.002 \\
5Y&	0.0017 & 	-0.0008 & 	0 & 	-0.0001 & 	0 & 	-0.0009 & 	0 & 	0.0001 & 	-0.0006 & 	 & 	0.0022 \\
&	 & 	 & 	 & 	 & 	 & 	 & 	 & 	 & 	 & 	 & 	 \\  \hline
$L^2$ norm&	0.0065 & 	0.0051 & 	0 & 	0.0007 & 	0 & 	0.0014 & 	0 & 	0.0029 & 	0.0073 & 	 & 	0.0115 \\
  \hline
\end{tabular}
\label{table:table.vol.discrepancies.vannavolga.epxiryandtenor.USDJPY}
\end{table}

\begin{table}
\caption{Discrepancies between the proposed volatilities and market volatilities, USDILS; 0.0001 corresponds to 0.01\% of implied volatility.}
\addtolength{\tabcolsep}{-2pt}
\begin{tabular}{|c	|l	|l	|l	|l	|l	|l	|l	|l	|l	|l	|l|}
  \hline
Expiry&	10put & 	15put & 	25put & 	35put & 	ATM & 	35call & 	25call & 	15call & 	10call & 	 & 	$L^2$ norm \\ \hline
2W&	0.0015&	0.0008&	0&	-0.0001&	0&	-0.0001&	0&	0.0010&	0.0021&	&	0.0029\\
3W&	0.0003&	0.0002&	0&	0.0001&	0&	0&	0&	0.0005&	0.0011&	&	0.0013\\
1M&	-0.0004&	-0.0002&	0&	0.0002&	0&	0.0001&	0&	0&	0.0002&	&	0.0005\\
2M&	-0.0002&	-0.0001&	0&	0.0001&	0&	0&	0&	0.0001&	0.0004&	&	0.0005\\
3M&	-0.0003&	-0.0002&	0&	0.0001&	0&	-0.0001&	0&	-0.0001&	0.0003&	&	0.0005\\
4M&	-0.0004&	-0.0003&	0&	0.0001&	0&	0&	0&	0.0002&	0.0007&	&	0.0009\\
6M&	-0.0008&	-0.0005&	0&	0.0002&	0&	0&	0&	-0.0004&	-0.0001&	&	0.0010\\
9M&	-0.0011&	-0.0007&	0&	0.0003&	0&	0&	0&	-0.0005&	-0.0004&	&	0.0015\\
1Y&	-0.0017&	-0.0011&	0&	0.0003&	0&	-0.0001&	0&	-0.0008&	-0.0008&	&	0.0023\\
18M&	-0.0004&	-0.0004&	0&	0.0003&	0&	0&	0&	0.0003&	0.0015&	&	0.0017\\
2Y&	-0.0002&	-0.0002&	0&	0.0004&	0&	-0.0002&	0&	-0.0002&	0.0007&	&	0.0009\\
3Y&	-0.0009&	-0.0006&	0&	0.0005&	0&	-0.0002&	0&	0.0004&	0.0019&	&	0.0023\\
4Y&	-0.0012&	-0.0007&	0&	0.0006&	0&	-0.0003&	0&	0.0008&	0.0028&	&	0.0033\\
5Y&	-0.0015&	-0.0009&	0&	0.0006&	0&	-0.0003&	0&	0.0008&	0.0026&	&	0.0033\\
&	&	&	&	&	&	&	&	&	&	&	\\ \hline
$L^2$ norm&	0.0034&	0.0022&	0&	0.0013&	0&	0.0006&	0&	0.0020&	0.0053&	&	0.0071\\
\hline
\end{tabular}
\label{table:table.vol.discrepancies.circles.epxiryandtenor.USDILS}
\end{table}

\begin{table}
\caption{Discrepancies between the vanna-volga volatilities and market volatilities, USDILS; 0.0001 corresponds to 0.01\% of implied volatility.}
\addtolength{\tabcolsep}{-2pt}
\begin{tabular}{|c	|l	|l	|l	|l	|l	|l	|l	|l	|l	|l	|l|}
  \hline
Expiry&	10put & 	15put & 	25put & 	35put & 	ATM & 	35call & 	25call & 	15call & 	10call & 	 & 	$L^2$ norm \\ \hline
2W&	-0.0004&	-0.0004&	0&	0.0001&	0&	0&	0&	-0.0001&	-0.0001&	&	0.0006\\
3W&	0.0008&	0.0002&	0&	-0.0001&	0&	-0.0001&	0&	0.0004&	0.0008&	&	0.0013\\
1M&	0.0014&	0.0006&	0&	-0.0002&	0&	-0.0002&	0&	0.0010&	0.0018&	&	0.0026\\
2M&	0.0015&	0.0006&	0&	-0.0001&	0&	-0.0001&	0&	0.0011&	0.0018&	&	0.0026\\
3M&	0.0015&	0.0006&	0&	-0.0001&	0&	-0.0001&	0&	0.0013&	0.0021&	&	0.0030\\
4M&	0.0019&	0.0008&	0&	-0.0001&	0&	-0.0002&	0&	0.0012&	0.0018&	&	0.0030\\
6M&	0.0024&	0.0011&	0&	-0.0002&	0&	-0.0002&	0&	0.0018&	0.0026&	&	0.0041\\
9M&	0.0028&	0.0013&	0&	-0.0002&	0&	-0.0003&	0&	0.0020&	0.0030&	&	0.0048\\
1Y&	0.0035&	0.0017&	0&	-0.0003&	0&	-0.0002&	0&	0.0022&	0.0028&	&	0.0053\\
18M&	0.0024&	0.0011&	0&	-0.0003&	0&	-0.0003&	0&	0.0013&	0.0005&	&	0.0030\\
2Y&	0.0022&	0.0009&	0&	-0.0003&	0&	-0.0002&	0&	0.0018&	0.0010&	&	0.0031\\
3Y&	0.0028&	0.0012&	0&	-0.0004&	0&	-0.0002&	0&	0.0011&	-0.0010&	&	0.0034\\
4Y&	0.0031&	0.0013&	0&	-0.0004&	0&	-0.0002&	0&	0.0007&	-0.0023&	&	0.0042\\
5Y&	0.0035&	0.0015&	0&	-0.0005&	0&	-0.0001&	0&	0.0006&	-0.0029&	&	0.0049\\
&	&	&	&	&	&	&	&	&	&	&	\\ \hline
$L^2$ norm&	0.0088&	0.0039&	0&	0.0010&	0&	0.0007&	0&	0.0049&	0.0074&	&	0.0131\\
  \hline
\end{tabular}
\label{table:table.vol.discrepancies.vannavolga.epxiryandtenor.USDILS}
\end{table}

\clearpage
\subsection{Distributions mostly supported on $\mathbb{R}_{+}$} %  beyond \(\mathbb{R}_{+}\)

\par Here examples of translated Student's $t$ distribution and normal distributions, both with very small $P(0)$, are considered. The densities are rescaled $p(X) \rightarrow \hat{p}(X) = \frac{p(X)}{1 - P(0)}$ so that $\hat{P}(\infty) - \hat{P}(0) = 1$ and only the part supported on $\mathbb{R}_{+}$ is further used for computing the KL divergence. As figures \ref{fig:example.pdf.student.smile} and \ref{fig:example.pdf.normal.smile} demonstrate, the distributions reconstructed from circular representations fitted to three points provide a fair approximation to the original distributions.

\begin{figure}[hbt!]% [h!]
\centering
\includegraphics[scale=0.85]{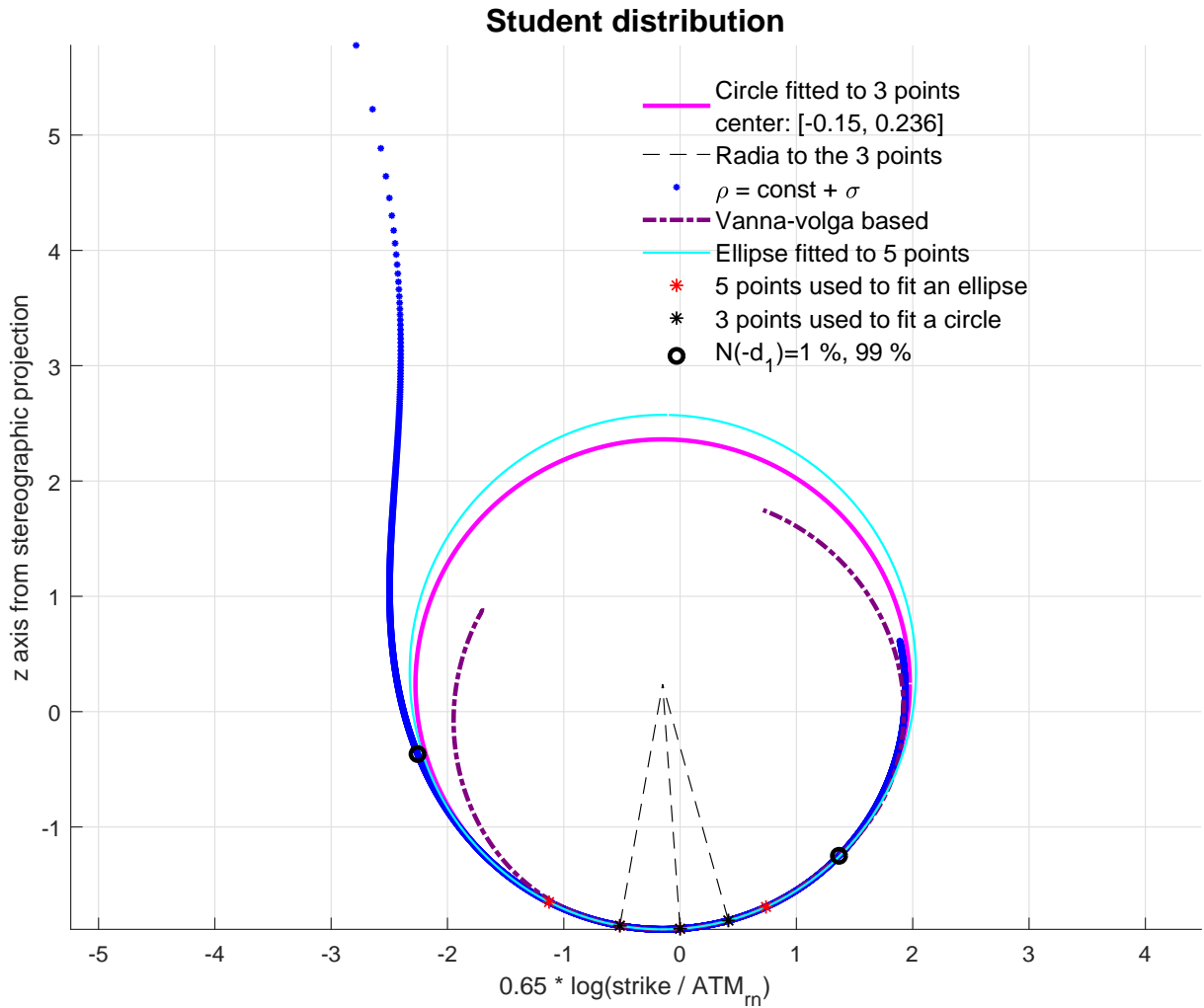}
\caption{Illustration of how a distribution characterized with its implied volatility profile can be approximated by inverting the circular representation. Geometric representation of translated Student's $t$ distribution with expectation $E = 3.7201$ and $\nu = 7.3824$ (blue asterisks) is approximated with a circle translated from the origin (magenta). The circle, in turn, defines another probability distribution that provides an approximation to represented translated Student's $t$ distribution as demonstrated in Figure \ref{fig:example.pdf.student.smile}.}
\label{fig:example.stereographic.student}
\end{figure}

% The same distribution, fit to PDF.
\begin{figure}[hbt!] % [h!]
\centering
\includegraphics[scale=0.9]{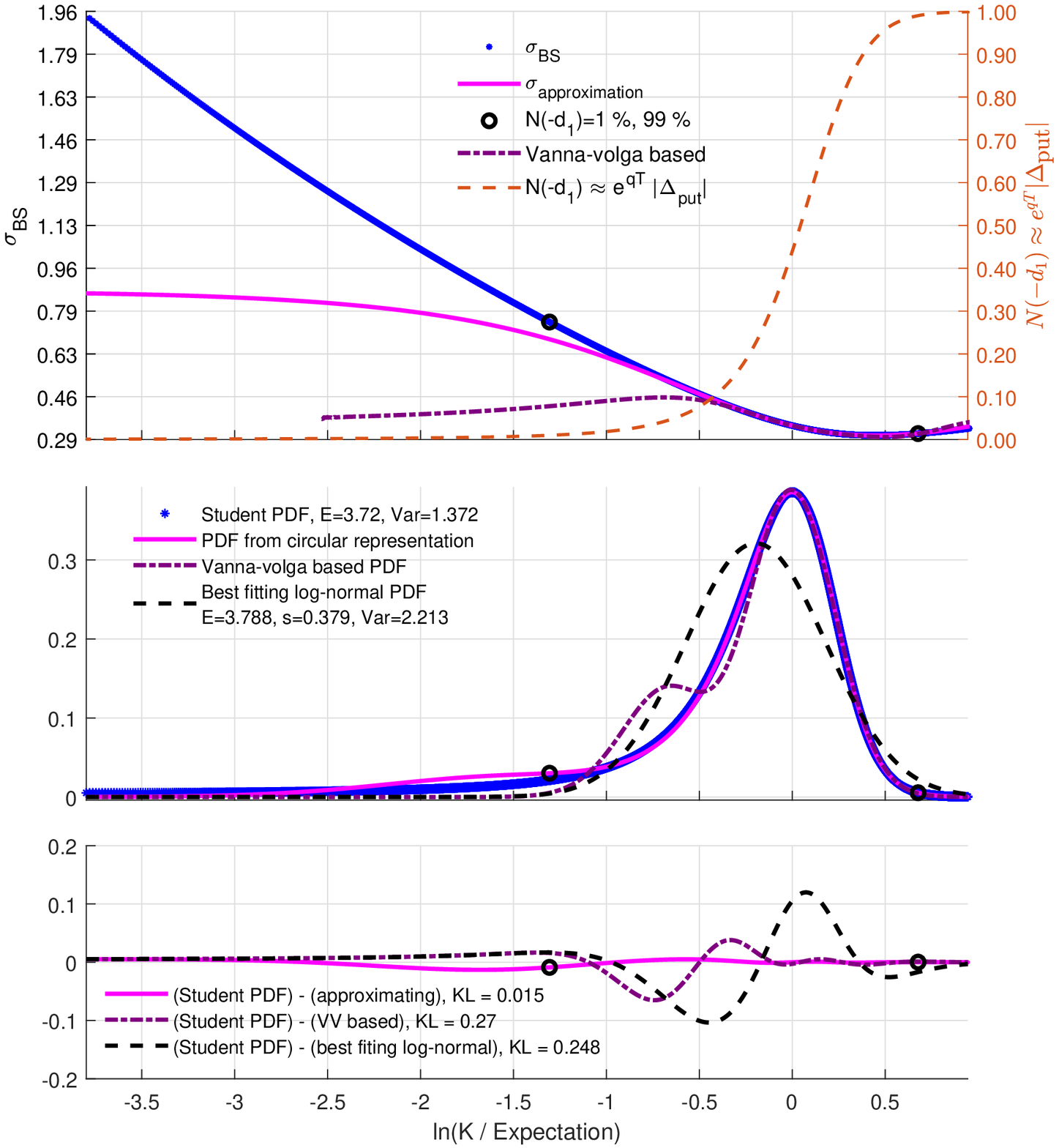}
\caption{Geometric representation of the translated Student's $t$ distribution from Figure \ref{fig:example.stereographic.student} and corresponding circular representation are further analyzed. \textbf{Upper part}: Volatility smile for the analyzed  distribution (blue), volatility smile derived from the circle fitted to the geometric representation (magenta) and implied volatility computed with the vanna-volga method (dashed violet). The two small black circles correspond to the strikes at which $N(-d_1) = 0.01,\, 0.99$.  \textbf{Middle part}: PDF of the translated Student's $t$ distribution (blue) and PDF whose geometric representation is circular approximation (magenta) are close one to another. Log-normal PDF that provides the best fit to analyzed PDF in the Kullback-Leibler sense is depicted with dashed blue. The values of Kullback-Leibler divergence between the distributions are presented in the legend. \textbf{Below part}: Difference between the analyzed PDF and (1) PDF represented with the circle (magenta), (2) vanna-volga based PDF (dashed violet) and (3) best fitting log-normal PDF (dashed black).}
\label{fig:example.pdf.student.smile}
\end{figure}

% The same distribution, geometric signature.
\begin{figure}[hbt!] % [h!]
\centering
\includegraphics[scale=0.7]{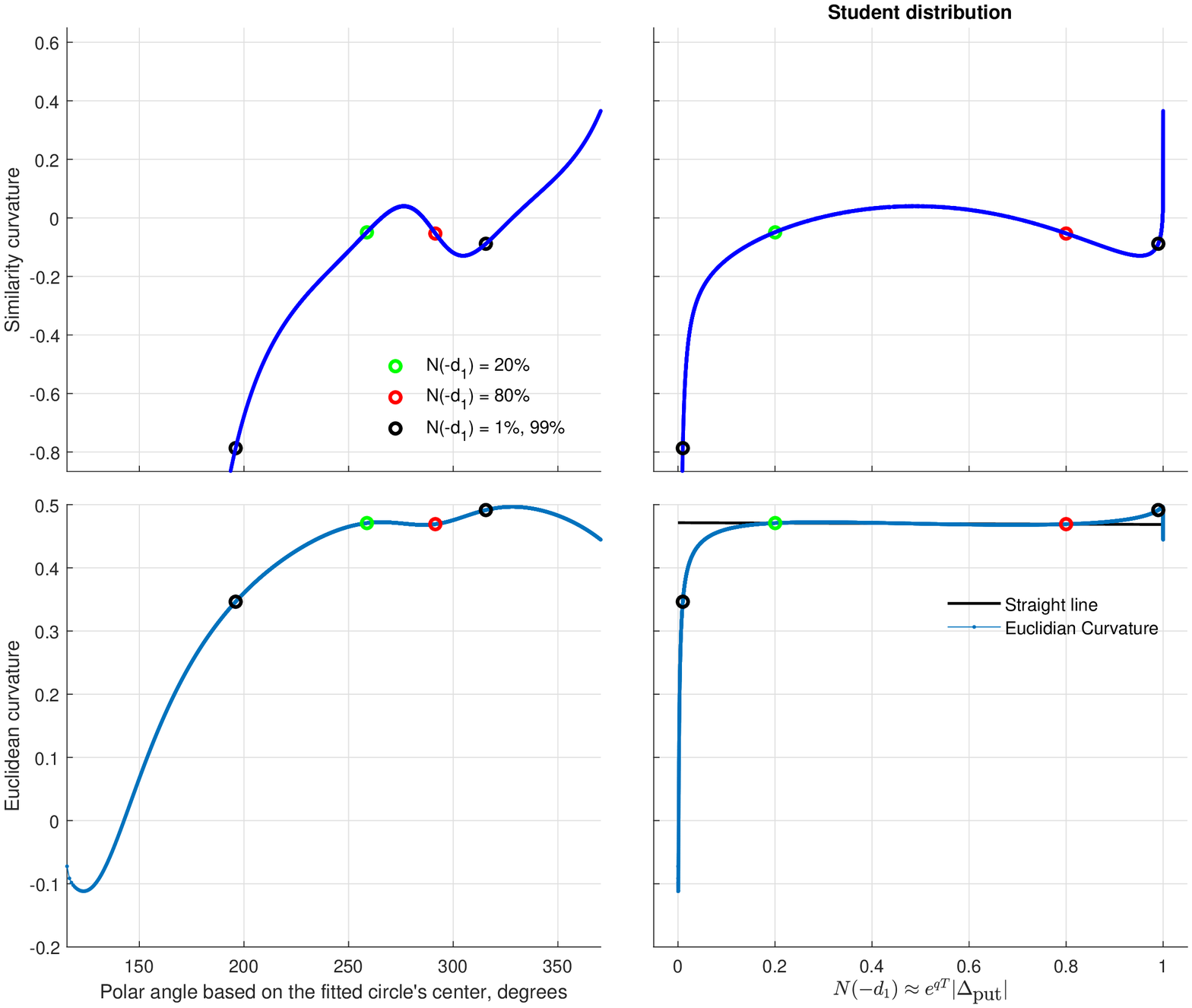}
\caption{Similarity and Euclidean curvatures of the geometric representation of the translated Student's $t$ distribution from Figure \ref{fig:example.stereographic.student}.}
\label{fig:example.pdf.student.geomsignature}
\end{figure}

\clearpage

%
% The figure was created using the function vols2sphere_different_distributions.m
% with gamma distribution.
% count_std_normal = 14
% count_smile = 8
%
%
%distributionS =
%
%  struct with fields:
%
%                       expectation: 11.3328
%                          gammastd: 2.6125
%                 Type_distribution: 'normal'
%                             gammA: []
%                        student_nu: 2.3433
%           How_to_accumulate_delta: 'differentiation_based_delta'
%                             theta: 0.6022
%                             kappa: 18.8178
%                   mixedlognormalS: [1×1 struct]
%                          variance: 9
%                           Strikes: [1×5001 double]
%    SIgmabs_from_givendistribution: [8×5001 double]

\begin{figure}[hbt!] %[h!]
\centering
\includegraphics[scale=0.9]{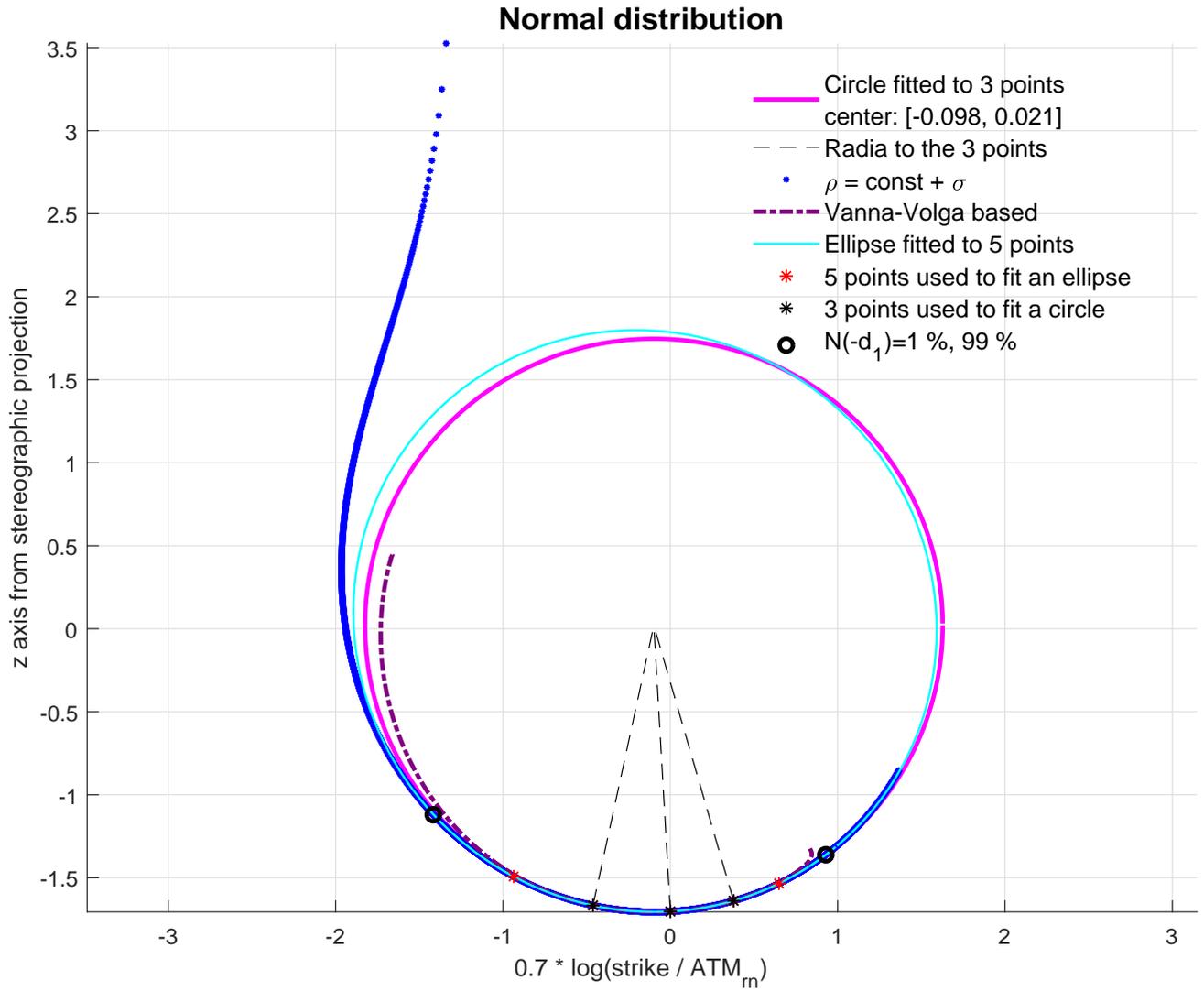}
\caption{Illustration of how a distribution characterized with its implied volatility profile can be approximated by inverting the circular representation. Geometric representation of normal distribution with $s = 3,\, \mu = 11.3328$ (blue asterisks) is approximated with a circle translated from the origin (magenta). The circle, in turn, defines another probability distribution that provides an approximation to the represented normal distribution as demonstrated in Figure \ref{fig:example.pdf.normal.smile}.}
\label{fig:example.circle.smile.normal}
\end{figure}

% The same distribution, fit to PDF.
\begin{figure}[hbt!] %[h!]
\centering
\includegraphics[scale=0.9]{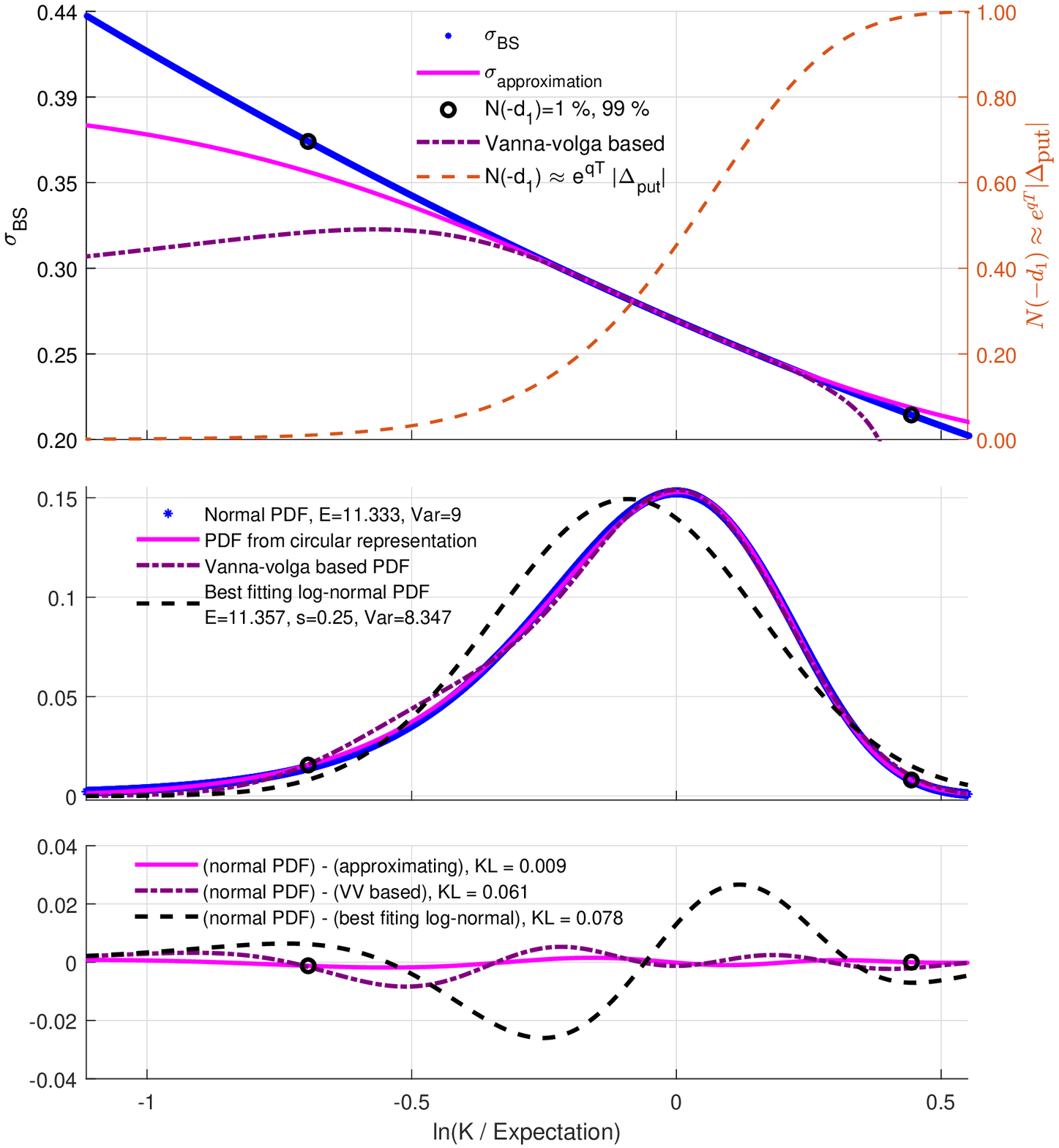} % Normal_PDF_fits_KLvals_countSTDnormal14_countSMILE8.eps}
\caption{Geometric representation of the normal distribution from Figure \ref{fig:example.circle.smile.normal} and corresponding circular representation are further analyzed. \textbf{Upper part}: Volatility smile for the considered normal distribution (blue), volatility smile derived from the circle fitted to the geometric representation (magenta) and implied volatility computed with the vanna-volga method (dashed violet). The two small black circles correspond to the strikes at which $N(-d_1) = 0.01,\, 0.99$.  \textbf{Middle part}: PDF of the translated Student's $t$ distribution (blue) and PDF whose geometric representation is circular approximation (magenta) are close one to another. Log-normal PDF that provides the best fit to analyzed normal PDF in the Kullback-Leibler sense is depicted with dashed blue. The values of Kullback-Leibler divergence between the distributions are presented in the legend. \textbf{Below part}: Difference between the analyzed PDF and (1) PDF represented with the circle (magenta), (2) vanna-volga based PDF (dashed violet) and (3) best fitting log-normal PDF (dashed black).}
\label{fig:example.pdf.normal.smile}
\end{figure}

% The same distribution, geometric signature.
\begin{figure}[hbt!] % [h!]
\centering
\includegraphics[scale=0.7]{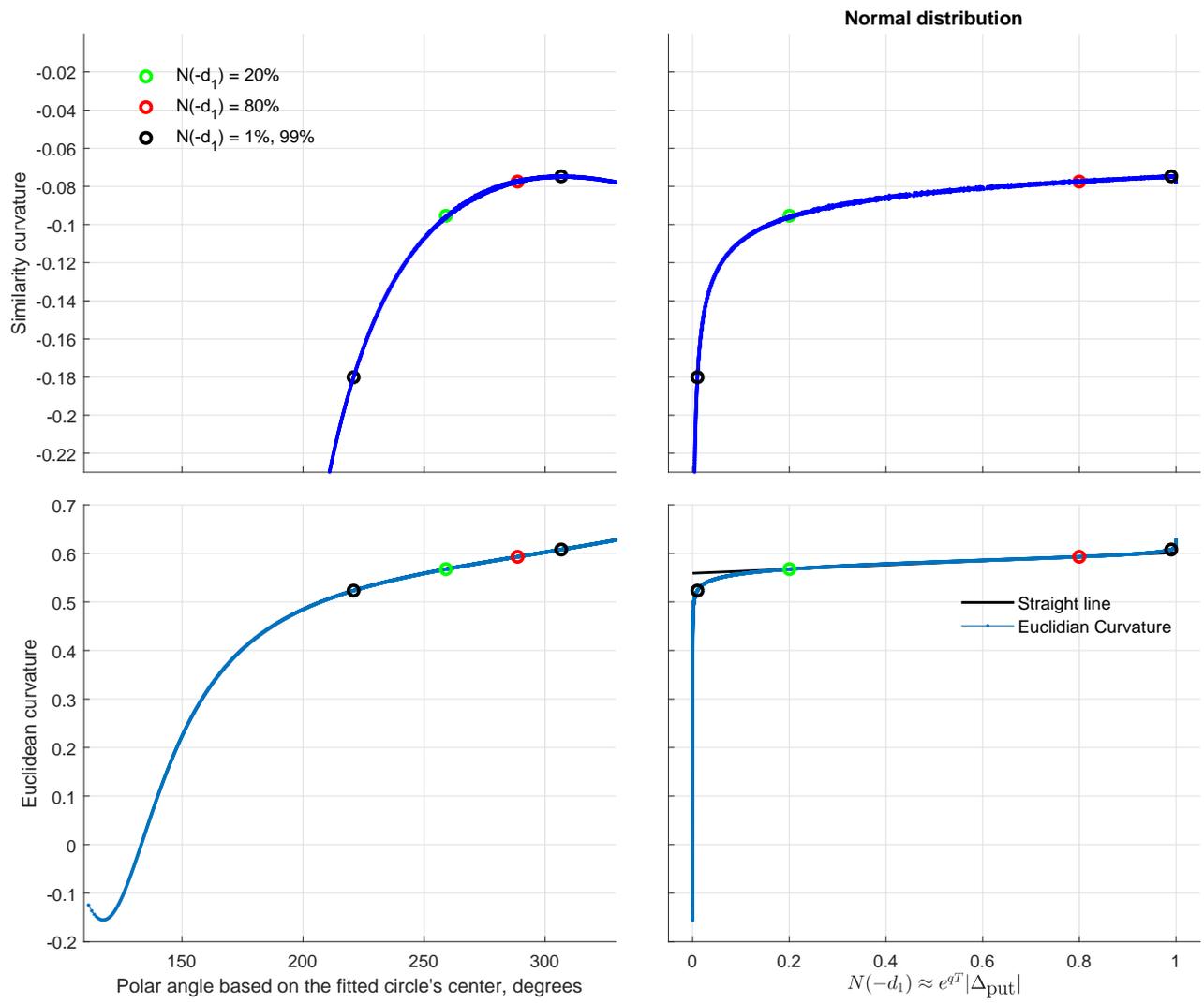}
\caption{Similarity and Euclidean curvatures of the geometric representation of the normal distribution from Figure \ref{fig:example.circle.smile.normal}.}
\label{fig:example.pdf.normal.geomsignature}
\end{figure}

\clearpage

\subsection{Negative probability}% \label{sec:Negative.probability}

\par
\begin{figure}[hbt!] %[h!]
\centering
\includegraphics[scale=0.85]{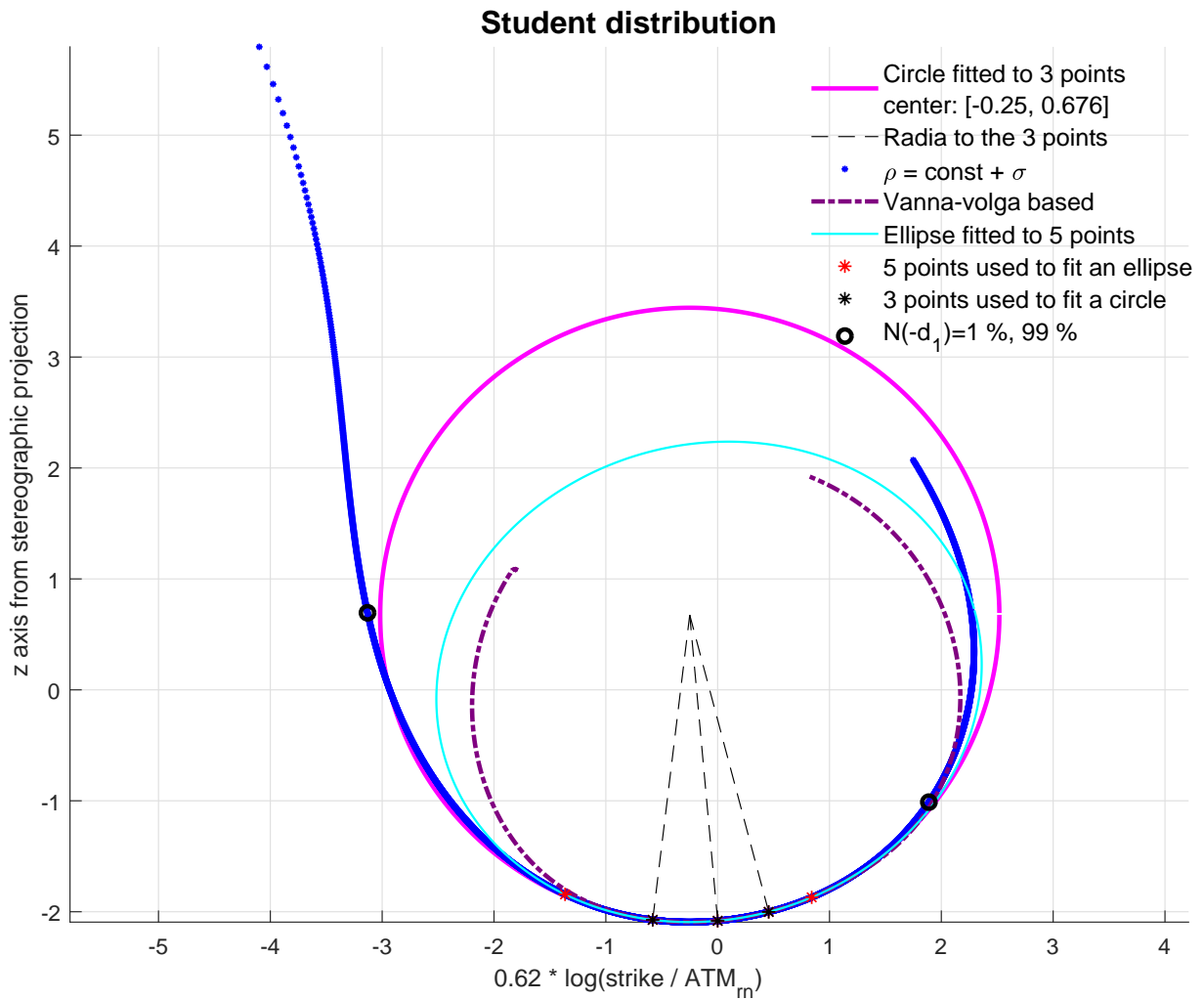}
\caption{Illustration of negative probability arising from inverting the circular representation. Geometric representation of translated Student's $t$ distribution with expectation $E = 3.7322$ and $\nu = 3.9565$ (blue asterisks) is approximated with a circle translated from the origin (magenta). The circle, in turn, defines another distribution and for the present case some of distribution's values are negative as demonstrated in Figure \ref{fig:example.smile.negativeprob.student}. Standard deviation of the analyzed Student's $t$ distribution is relatively close to its expectation that causes the implied volatility described by the fitted circle to underlie negative probabilities.}
\label{fig:example.stereographic.negativeprob.student}
\end{figure}

\begin{figure}[hbt!] %[h!]
\centering
\includegraphics[scale=0.9]{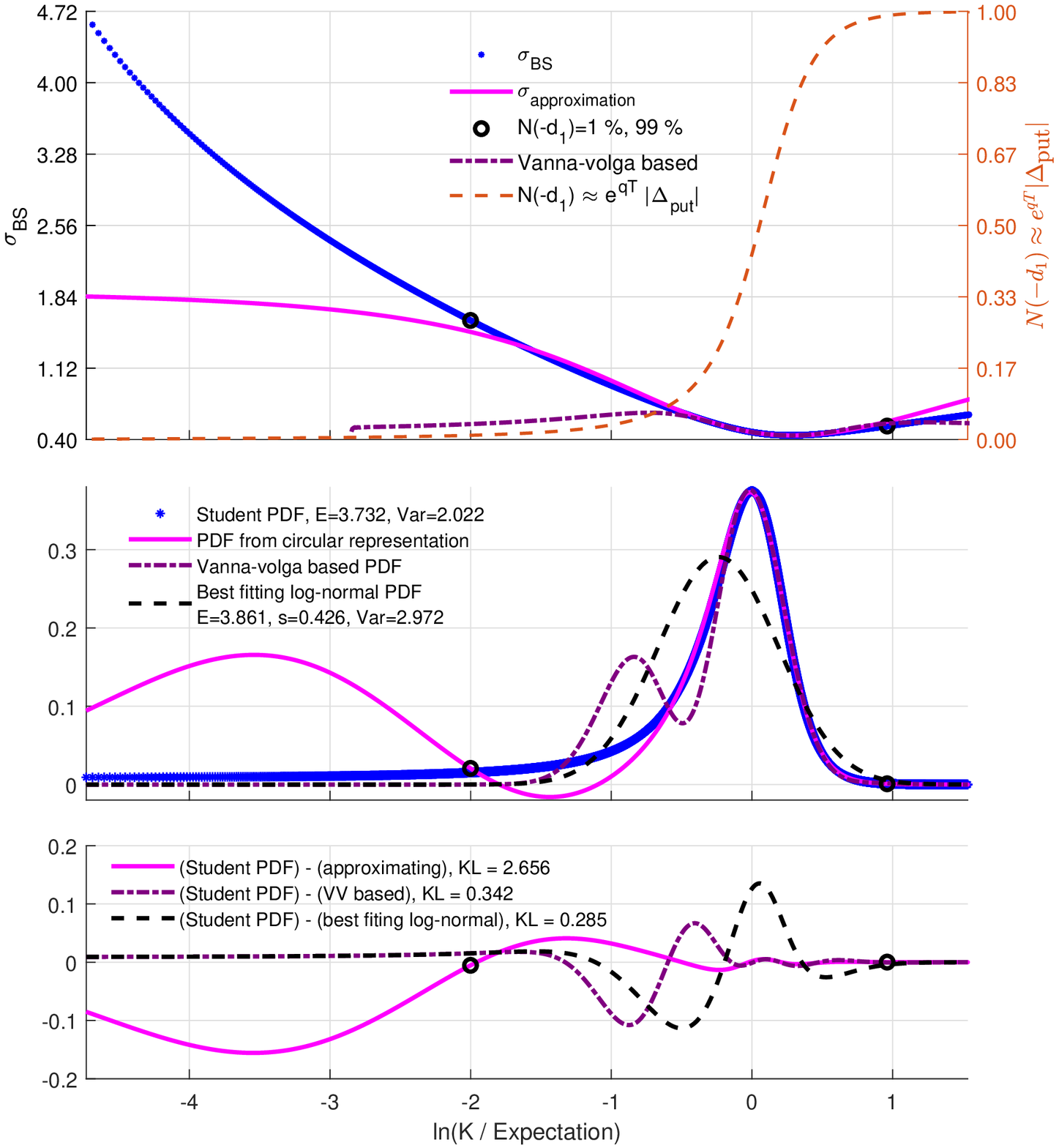} % Normal_PDF_fits_KLvals_countSTDnormal14_countSMILE8.eps}
\caption{Geometric representation of the translated Student's $t$ distribution from Figure \ref{fig:example.stereographic.negativeprob.student} and corresponding circular representation are further analyzed. \textbf{Upper part}: Volatility smile for the analyzed  distribution (blue), volatility smile derived from the circle fitted to the geometric representation (magenta) and implied volatility computed with the vanna-volga method (dashed violet). The two small black circles correspond to the strikes at which $N(-d_1) = 0.01,\, 0.99$.  \textbf{Middle part}: PDF of the translated Student's $t$ distribution (blue) and PDF whose geometric representation is circular approximation (magenta) are close one to another. Log-normal PDF that provides the best fit to analyzed PDF in the Kullback-Leibler sense is depicted with dashed blue. The values of Kullback-Leibler divergence between the distributions are presented in the legend. \textbf{Below part}: Difference between the analyzed PDF and (1) PDF represented with the circle (magenta), (2) vanna-volga based PDF (dashed violet) and (3) best fitting log-normal PDF (dashed black).}
\label{fig:example.smile.negativeprob.student}
\end{figure}

 Negative density values may arise by inverting shapes known in the representation space as demonstrated below. The KL divergence is not defined for negative distribution values. To be able to apply the formula for computing the KL divergence, negative values of $p(x)$ were replaced with very small positive ones (e.g. of the order of $10^{-50}$) and the density was rescaled to achieve $\hat{P}(\infty) = 1$, then the values of pseudo KL divergence were computed in such cases. An example of a circle that represents density obtaining negative values is demonstrated in Figure \ref{fig:example.stereographic.negativeprob.student}. The pseudo KL divergence between specific translated Student's $t$ distribution and the density represented by the circle fitted to corresponding geometric representation is relatively large as can be seen in Figure \ref{fig:example.smile.negativeprob.student}. The considered Student's $t$ distribution has non-negligible part supported outside $\mathbb{R}_{+}$ as its standard deviation is not small enough versus the expectation. This causes the abnormality with negative density underlying the fitted circle in the representation space.
%
% The figure was created using the function vols2sphere_different_distributions.m
% with gamma distribution.
% count_std_normal = 13
% count_smile = 7
%
%
%distributionS =
%
%  struct with fields:
%
%                       expectation: 3.7322
%                          gammastd: 1.4220
%                 Type_distribution: 'Student'
%                             gammA: []
%                        student_nu: 3.9565
%           How_to_accumulate_delta: 'differentiation_based_delta'
%                             theta: 0.5418
%                             kappa: 6.8882
%                   mixedlognormalS: [1×1 struct]
%                          variance: 2.0222
%                           Strikes: [1×5001 double]
%    SIgmabs_from_givendistribution: [7×5001 double]

% Negative probability

% The same distribution, fit to PDF.

% The same distribution, geometric signature.
\begin{figure}[hbt!] % [h!]
\centering
\includegraphics[scale=0.7]{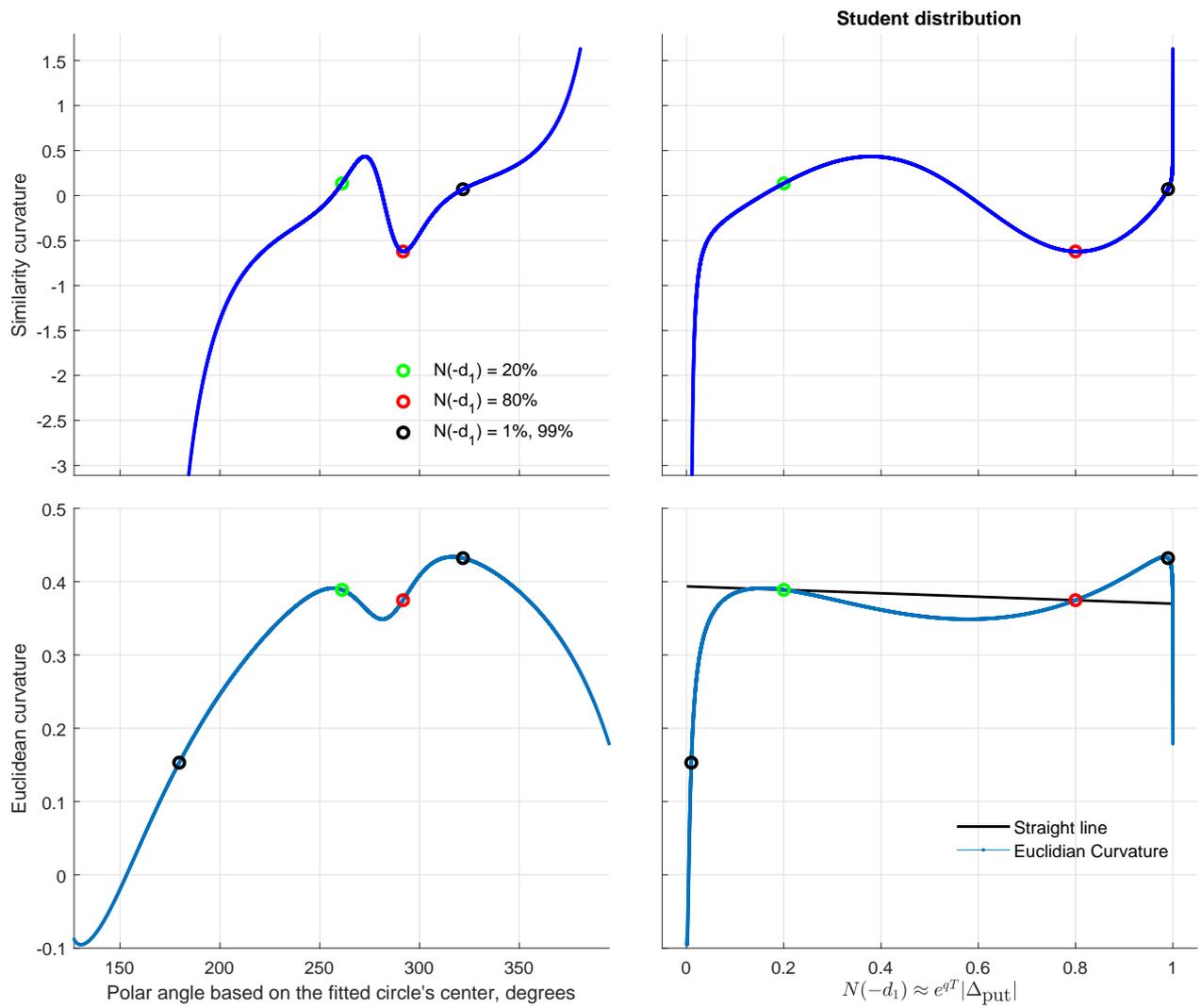}
\caption{Similarity and Euclidean curvatures of the geometric representation of the translated Student's $t$ distribution from Figure \ref{fig:example.stereographic.negativeprob.student}.}
\label{fig:example.pdf.negativeprob.geomsignature}
\end{figure}

\clearpage

\section{Discussion}
\par A wide range of cognitive processes may utilize a rich machinery for geometric computations in the brain \cite{Polyakov:2019}. Does the brain incorporate its geometric intuition into its probabilistic reasoning? And if yes to what extent and how? On the practical side, FX options' market makers provide pricing of uncertainty based on their experience and gut feeling, and measure uncertainty in terms of implied volatility. This work establishes a methodology of how to geometrically represent uncertainty measured with implied volatility in a way that allows to complete the knowledge about the represented probability distribution by extending the measurement of implied volatility to arbitrary strikes from only a small number of known values. The methodology also allows visualization of probability distributions based on intuitive geometric symmetries. For example, log-normal distributions are represented with circles centered at the origin and circles not centered at the origin seem to represent probability densities that closely approximate densities with bell-shaped profile.
\par The ``Results'' section presents examples of geometric representation for a number of distributions with both bell-shaped and non-bell-shaped probability density profiles. Geometric representations of probability densities $\{q(x)\}$ with bell-shaped profile, fully or mostly supported on $\mathbb{R}_{+}$, were successfully approximated with circles; those circles, in turn, represent feasible probability distributions $\bar{q}(x)$ (reconstructed density is non-negative: $\bar{q} \geq 0$). Those reconstructed densities $\bar{q}(x)$ provide good approximation to the original distributions $q(x)$ as measured with the Kullback-Leibler divergence. The above mentioned approximability based on circles in the representation domain is typical for the analyzed probabilities with bell-shaped density profiles. The circles-based approximability does not hold for probabilities whose density is non-bell-shaped, for example for uniform distributions. It may not hold for the densities whose cumulative probability $P(0)$ (support outside of $\mathbb{R}_{+}$) is non-negligible as demonstrated in Figure \ref{fig:example.smile.negativeprob.student}. 
\par For all considered distributions with bell-shaped densities, their approximations with distributions arising from circular representations were noticeably superior to the approximations based on the vanna-volga method (Figures \ref{fig:example.pdf.gamma.smile}, \ref{fig:example.pdf.student.smile}, \ref{fig:example.pdf.normal.smile}). The approximations of considered market volatility smiles were in general better for circular representations as well (Tables \ref{table:table.vol.discrepancies.circles.epxiryandtenor.EURUSD} - \ref{table:table.vol.discrepancies.vannavolga.epxiryandtenor.USDILS}). So, the proposed method looks more precise for extending the volatility smile to entire continuum of strikes versus the vanna-volga approach, though circular approximation has one free parameter while the vanna-volga method is parameter-free. Ellipses have also been fit to the geometric representations based on five points that define a unique ellipse. For considered examples, in general, the fit with ellipses does not seem to be superior over the fit with circles while only three points are needed to pass a circle. Implied volatility contains information that, for bell-shaped densities at least, may allow completion based on just three values; what is not the case for point-wise probability values. While use of implied volatility and ``delta'' (see Table \ref{tab:meaning.variable.callprice}) in finance is convenient as it allows to apply a kind of uniform measure to strikes of options with different maturities, this technical convenience does not contradict the biological rationale introduced in the present work.
\par Geometric representation of probabilities allows to introduce geometric equivalence classes for distributions. In particular, circular representations are in the focus of the present work and they presumably either underlie densities with bell-shaped profile or approximate their representations. Real-world applications often deal with approximate data due to bid-ask spreads in finance and perception/action (measurement/implementation) imperfections in general. Therefore, mathematically defined class of geometric equivalence may be enriched with good enough approximations for many practical purposes and this should be accounted for. Specifically, biological systems may not need to distinguish small drifts in the represented probability distributions and so they may rely on equivalence classes that allow approximating geometric shapes in the representation space. So, accounting for small imperfections, circular geometric representations may actually cover representation of probability densities with bell-shaped profiles that are supported on $\mathbb{R}_{+}$; this claim needs to get stronger support yet.
\par For probability densities $p(X)$ supported outside of $\mathbb{R}_{+}$ geometric representation can be based on the probabilities supported strictly on $\mathbb{R}_{+}$ by applying one-to-one transformations, for example
\begin{equation}\label{eq:exponential.transformation}
    p(x) \rightarrow \hat{p}(\hat{x}),\; \hat{x} = e^{x} \; .
\end{equation}
\par Distributions defined by implied volatility like in formula \eqref{eq:p.throughDDcallDK2} depend on market characteristics $r$, $q$, $T$, $S_0$ from the Black-Scholes-Merton formula \eqref{eq:BSM.call} and this ``market-specificity'' should be accounted for while inverting geometric representation into the corresponding probability distribution. Construction of geometric representation for a given probability distribution is ``market-specific'' in the same manner, so that the values of $S_0$, $T$, $r$, $q$ should be introduced. Those ``market'' variables are not disconnected from the distribution as any implied distribution $p(x)$ imposes constraint on them based on equation \eqref{eq:S0.expectation}:
  $\mathbb{E}_{p} = S_0 e^{(r - q) T} $ and vice versa, the reconstructed probability distribution should satisfy the above constraint.
\par Furthermore, any non-negative integrable function $f$ defined on $\mathbb{R}$ can be represented with a curve in polar coordinate system by transforming $f$ into a probability density \footnote{Function $f$ should be rescaled with constant $c$ to achieve $\int (c \cdot f(x)) dx = 1 $ and transformation \eqref{eq:exponential.transformation} or another relevant transformation should be applied if necessary to adjust the support.} and further by representing the resulting probability distribution with the procedure introduced in this work.

\par Family of normal distributions is closed under convolution transformations while the set of log-normal distributions is not. So, in order to keep distributions represented with circles centered at the origin closed under the convolution, their argument should be log-transformed and the inverse transformation  \eqref{eq:exponential.transformation} should be applied to the result of the convolution. Consider the following open questions and problems related to the proposed concept of geometric representation of probability:
\begin{enumerate}
    \item What operations and for what classes of distributions preserve (or approximately preserve) geometric properties of the underlying geometric representations\footnote{That is are invariant.}, like circles being mapped to circles (or curves approximating circles well enough) and so on?
    \item How precisely do distributions represented with circles approximate arbitrary distributions with bell-shaped profile, what would be quantitative conditions on those shapes? 
    \item What are possible equivalence classes for probability distributions, from the point of view of geometric representation, allowing also for small drifts, and how those classes can be used in applications and research?
    \item Finding methods for efficient completion of probabilities in general. Here completion of probabilities with bell-shaped densities was implemented based on circular representations.
\end{enumerate}

\appendix

\section{Probability density expressed through implied volatility}

%%% For Appendix
% format the equation environment
%\renewcommand{\theequation}{A.\arabic{equation}}
\numberwithin{equation}{section}
%\renewcommand{\theequation}{\thesection.\arabic{equation}}
% reset the counter
\setcounter{equation}{0}

\par Here equations \eqref{eq:2nd.derivative.call} and \eqref{eq:BSM.call} are used to derive a differential expression based on $\sigma(K)$ for probability density $p$. Equation \ref{eq:2nd.derivative.call} implies

\begin{equation}\label{eq:preliminary.p.throughDDcallDK2}
    p(K) = e^{r T} \frac{\partial^2 \mbox{call}(K)}{\partial K^2} =   e^{r T}
        \mbox{call}''(K)\,.
\end{equation}

Now differentiate the expression for $\mbox{call}(K)$ from equation \eqref{eq:BSM.call}:
\begin{equation*}
    \mbox{call}' \equiv \displaystyle\frac{\partial \mbox{call}(K)}{\partial K} = S_0 e^{-q T} n(d_1) \frac{\partial d_1}{\partial K} - e^{-r T} N(d_2) - K e^{-r T} n(d_2)  \frac{\partial d_2}{\partial K}
\end{equation*}
and apply \eqref{eq:lemma.equilibrium} to get
\begin{equation}\label{eq:dcallDK.with.sigma.of.K}
    \mbox{call}'(K) = S_0 e^{-q T} n(d_1) \cdot  \frac{\partial }{\partial K} (d_1 - d_2) - e^{-r T} N(d_2) = S_0 e^{-q T} n(d_1) \sigma' \sqrt{T}  -e^{-r T} N(d_2) \, .
\end{equation}
Differentiation of \eqref{eq:dcallDK.with.sigma.of.K} with respect to $K$ leads to the expression:
\begin{equation}\label{eq:ddcallDDK.with.sigma.of.K}
    \mbox{call}''(K) = \frac{1}{K  \sigma \sqrt{2 \pi} \sqrt{T}} e^{-r T - \frac{{d_2}^2}{2}}  \left[ 1 + 2 K \sqrt{T} d_1 \sigma' + K^2 T \left(d_1 d_2 {\sigma'}^2 + \sigma \sigma'' \right) \right] \, .
\end{equation}
Now, formulae \eqref{eq:preliminary.p.throughDDcallDK2}, \eqref{eq:ddcallDDK.with.sigma.of.K} imply
\begin{equation}\label{eq:p.throughDDcallDK2}
    p(K) =  \frac{ 1 + 2 K \sqrt{T} d_1 \sigma' + K^2 T \left(d_1 d_2 {\sigma'}^2 + \sigma \sigma'' \right)}{K  \sigma \sqrt{2 \pi} \sqrt{T}} e^{- \frac{{d_2}^2}{2}} \, .
\end{equation}
Whenever an approximate geometric representation for probability  is known (meaning that corresponding $\sigma(K)$ is known too) and has a functional form, for example a circle, a function $\sigma(K)$ can be used in equation \eqref{eq:p.throughDDcallDK2} to obtain an approximate probability density function. If an approximate geometric representation has no known functional form but is available point-wise, equation \eqref{eq:p.throughDDcallDK2} may still be useful in the form of finite difference for dense enough points. Approximate geometric representation may lead to negative values of probability density $p(K)$ for some $K$.
Equation \eqref{eq:p.throughDDcallDK2} implies condition on $\sigma(K)$ that is equivalent to non-negativity of probability density:
\begin{equation}\nonumber%\label{eq:condition.nonnegative.probability}
    1 + 2 K \sqrt{T} d_1 \sigma' + K^2 T \left(d_1 d_2 {\sigma'}^2 + \sigma \sigma'' \right) \geq 0 \, .
\end{equation}
\par Denote differentiation with respect to $\ln K$ with dot:
$$\dot{\sigma} \equiv \frac{d \sigma}{ d (\ln K)} $$
to get
\begin{equation}\nonumber %\label{eq:p.throughDDcallDLogK2}
    p(K) =  \frac{1 + \sqrt{T} \left( d_1 + d_2 \right)  \dot{\sigma} + T d_1 d_2  {\dot{\sigma}}^2 + T \sigma \ddot{\sigma}}{K  \sigma \sqrt{2 \pi} \sqrt{T}} e^{- \frac{{d_2}^2}{2}} \, .
%    p(K) =  \frac{\left( 1 + \sqrt{T} d_1 \dot{\sigma} \right) \left( 1 + \sqrt{T} d_1 \dot{\sigma} - T \sigma \dot{\sigma} \right) + T \sigma \ddot{\sigma}}{K  \sigma T \sqrt{2 \pi}} e^{- \frac{{d_2}^2}{2}} \, .
\end{equation}

\bibliographystyle{ieeetr} %unsrt} %
\bibliography{xbib_all}

\end{document}